\documentclass[lettersize,journal]{IEEEtran}
\usepackage{amssymb}
\usepackage{amsmath,amsfonts}
\usepackage{algorithmic}
\usepackage{algorithm}
\usepackage{array}
\usepackage[caption=false,font=normalsize,labelfont=sf,textfont=sf]{subfig}
\usepackage{textcomp}
\usepackage{stfloats}
\usepackage{url}
\usepackage{verbatim}
\usepackage{graphicx}
\usepackage{cite}
\usepackage{xcolor}  %
\usepackage{stfloats}
\usepackage{bm}
\usepackage{booktabs}
\usepackage{tabularx}
\usepackage{colortbl}  
\usepackage{subcaption}
 \DeclareMathSizes{10}{10}{5}{5}
\begin{document}

\title{Sensing-enabled Secure Rotatable Array System Enhanced by Multi-Layer Transmitting RIS}

\author{Maolin Li, Feng Shu, Minghao Chen, Cunhua Pan, Fuhui Zhou, Yongpeng Wu, Liang Yang
	% <-this % stops a space
	\thanks{This work was supported in part by the National Natural Science Foundation of China under Grant U22A2002, and by the Hainan Province Science and Technology Special Fund under Grant ZDYF2024GXJS292; in part by the Scientific Research Fund Project of Hainan University under Grant KYQD(ZR)-21008; in part by the Collaborative Innovation Center of Information Technology, Hainan University, under Grant XTCX2022XXC07; in part by the National Key Research and Development Program of China under Grant 2023YFF0612900. (Corresponding author: Feng Shu.)}% <-this % stops a space
	\thanks{Maolin Li and Minghao Chen are with the School of Information and Communication Engineering, Hainan University, Haikou 570228, China (e-mail: limaolin0302@163.com; 996739@hainanu.edu.cn).}
	\thanks{Feng Shu is with the School of Information and Communication Engineering, Hainan University, Haikou 570228, China, and also with the School of Electronic and Optical Engineering, Nanjing University of Science and Technology, Nanjing 210094, China (e-mail: shufeng0101@163.com).}
\thanks{Cunhua Pan is with the National Mobile Communications Research Laboratory, Southeast University, China (e-mail: cpan@seu.edu.cn).}
\thanks{Fuhui Zhou is with the College of Artiffcial Intelligence, Nanjing University
of Aeronautics and Astronautics, Nanjing 210016, China (e-mail:
zhoufuhui@ieee.org).}
\thanks{Yongpeng Wu is with the Shanghai Key Laboratory of Navigation and Location Based Services, Shanghai Jiao Tong University, Minhang, Shanghai, 200240, China (e-mail: yongpeng.wu2016@gmail.com).}
\thanks{Liang Yang is with the College of Computer Science and Electronic Engineering, Hunan University, Changsha 410082, China (e-mail: liangy@hnu.edu.cn). }
}
%\author{IEEE Publication Technologu,~\IEEEmembership{Staff,~IEEE,}
%        % <-this % stops a space
%\thanks{This paper was produced by the IEEE Publication Technology Group. They are in Piscatawau, NJ.}% <-this % stops a space
%\thanks{}}
%
%% The paper headers

\maketitle

\begin{abstract}
  Programmable metasurfaces and adjustable antennas are promising technologies. The security of a rotatable array system is investigated in this paper. A dual-base-station (BS) architecture is adopted, in which the BSs collaboratively perform integrated sensing of the eavesdropper (the target) and communication tasks. To address the security challenge when the sensing target is located on the main communication link, the problem of maximizing the secrecy rate (SR) under sensing signal-to-interference-plus-noise ratio requirements and discrete constraints is formulated. This problem involves the joint optimization of the array pose, the antenna distribution on the array surface, the multi-layer transmitting RIS phase matrices, and the beamforming matrices, which is non-convex. To solve this challenge, an two-stage online algorithm based on the generalized Rayleigh quotient and an offline algorithm based on the Multi-Agent Deep Deterministic Policy Gradient are proposed. Simulation results validate the effectiveness of the proposed algorithms. Compared to conventional schemes without array pose adjustment, the proposed approach achieves approximately 22\% improvement in SR. Furthermore, array rotation provides higher performance gains than position changes.
\end{abstract}


\section{Introduction}
\IEEEPARstart{M}assive multiple-input multiple-output (MIMO) technology has been extensively studied. With the massive connectivity of terminals, substantial energy consumption, high cost, and security concerns have emerged as critical challenges. To address these issues, programmable metasurfaces (PMs) composed of a large number of low-cost elements are regarded as a key technology for 6G and beyond networks~\cite{Danufane2021,Liu2021,DiRenzo2020}. Examples include stacked intelligent metasurfaces (SIMs)~\cite{An2025a}, flexible intelligent metasurfaces (FIMs)~\cite{An2025}, fluid antennas (FAs)~\cite{Wong2021}, movable antennas (MAs)~\cite{Zhu2024a}, and rotatable antennas (RAs)~\cite{Zhang2025}. By exploiting the spatial degrees of freedom (DoF), more efficient and reliable communications can be achieved, while also supporting other functionalities such as sensing and computing.

\subsection{Prior work}
Reconfigurable intelligent surfaces (RISs) are composed of a large number of small elements, the amplitude and phase shift of which can be adjusted to adapt to the communication environment, thereby enhancing system performance~\cite{Wang2025a}. Depending on whether the amplitude can be adjusted, RISs can be categorized into active RIS and passive RIS. A key challenge in RIS technology is the multiplicative fading effect. Active RIS, which offers higher compensation capability, has attracted increasing attention and achieves optimal performance when deployed near the base station (BS) or users~\cite{Wu2018,Kumar2025}. The flexible deployment and programmable capability of RIS have spurred research on novel types of metasurfaces. In~\cite{Mu2022}, a simultaneously transmitting and reflecting RIS (STAR-RIS) was investigated~\cite{Xue2025}, which significantly reduces power consumption compared to conventional purely reflecting or transmitting RISs. The secrecy performance of STAR-RIS assisted wireless communication systems was investigated in~\cite{Xiao2024}, where the semi-definite relaxation and Dinkelbach's algorithm were employed to achieve superior covert rates compared to conventional RIS. In practical implementations, the adjustable phase shifts of RISs exhibit limited resolution. Inspired by deep neural networks, multi-layer RIS architectures have been proposed to harness programmable interactions between different layers, aiming to unlock additional degrees of freedom for enhanced system performance~\cite{An2023}. %The study in demonstrated that, compared to conventional RIS-assisted MIMO systems, multi-layer RISs can achieve a $1.5$-fold capacity improvement while reducing hardware costs.
 In~\cite{Kavianinia2025}, the secrecy communication in a SIM-assisted system was investigated, demonstrating that increasing the number of metasurface layers and elements can effectively reduce interference and enhance the secrecy rate (SR) performance. However, an increased number of layers and elements implies higher hardware costs and system complexity. To address these limitations, several advanced metasurface architectures have been explored, including the intelligent omni-surfaces, beyond-diagonal RISs, FIMs, and holographic metasurface. These designs aim to overcome the hardware and signal processing constraints of conventional RIS by exploiting additional DoF to achieve higher theoretical performance bounds.

Most RIS technologies are investigated based on fixed element positions, including fixed antenna placements at the BS and static element configurations on the RIS. However, in scenarios such as multi-user communications or secure transmission, fixed antenna arrangements may struggle to guarantee satisfactory performance for users in disadvantaged locations. To address this issue, fluid/movable antennas enable positional adjustments through mechanical actuation~\cite{Huangfu2025}, and RA driven by hybrid mechanical and electronic controls can provide high responsiveness and be readily integrated into existing systems. By empowering dynamic tuning of antenna placement and beam steering, these technologies are regarded as promising candidates.

Antenna position optimization is crucial for MA technologies. However, driving antenna movement consumes additional power and introduces time delays~\cite{Shao2025b}. Under these practical constraints, effectively leveraging antenna mobility to achieve performance improvements while ensuring secure communication remains a key challenge. Reference \cite{Ma2024} demonstrated that MA arrays achieve superior sensing performance compared to conventional uniform planar arrays. In~\cite{Zhu2024}, with continuously movable antennas considered, the required transmit power was shown to be lower compared to conventional fixed-position antennas (FPAs). The problem of maximizing sensing performance under user minimum service guarantee was investigated in~\cite{Jiang2025}, demonstrating more efficient resource utilization compared to FPA systems. To reduce hardware cost and complexity, the position and rotation of antenna arrays have been investigated. Compared to FPA systems, six-dimensional (6D) MAs can achieve higher average sum rates in scenarios with non-uniform spatial distribution of users~\cite{Shao2025a}. Then, the study in~\cite{Shao2025} investigated the rate maximization problem under discrete positions and rotations, demonstrating that 6D MAs can effectively enhance network capacity~\cite{Pi2025,Zheng2025a,Zheng2025b}. The key challenge for RAs lies in optimizing the pointing vectors of their antenna elements. While most FPA and MA systems are studied under the assumption of omnidirectional antennas, users located in side lobes may experience significant signal attenuation and interference. To address this,~\cite{Zhang2025} investigated a RA system equipped with directional antennas, deriving an asymptotically optimal hybrid precoding scheme that demonstrates superior performance over systems with fixed antenna orientations. Furthermore, flexible and low-cost pinching antenna systems have been investigated, demonstrating improved transmission rates and enhanced security performance compared to FPA systems~\cite{Zhu2025,Lv2025}.

As mentioned above, various PMs have been extensively investigated for their potential to enhance security and system capacity. However, artificial noise (AN), as a key security-enabling technique, faces the practical limitation of high power consumption.~\cite{Shu2024}. This issue is particularly pronounced in the context of integrated sensing and communication (ISAC) and massive connectivity~\cite{Liao2024,Zhang2025b}, where BSs are required to simultaneously transmit radar signals, communication signals, and AN. A single BS may be unable to handle all these tasks effectively, thus failing to ensure communication security. To achieve high-precision long-range sensing, a collaborative detection framework between a UAV-aided aerial ISAC BS and a terrestrial BS was investigated in \cite{Wang2025,Zhang2025a}. While maintaining required sensing performance, the communication data rate was effectively enhanced. To mitigate the impact of path loss on both communication and sensing performance, the work in~\cite{Meng2025} achieved an effective trade-off between these two objectives by optimizing the size of the cooperative base station cluster and the transmit power.
%To address this challenge, a novel secure transmission network architecture is investigated in this paper.

\subsection{Our contributions}
Nevertheless, most technologies for PMs and adjustable antennas are predicated on isotropic antennas. When directional antennas are employed for secure communications, security issues remain, particularly when an eavesdropper (Eve) is located within the mainlobe of the communication. Effectively degrading the eavesdropper's demodulation capability with limited power poses a key challenge.
To address this challenge, a novel multi-layer RIS-enhanced dual-BS secure communication network is investigated. The main contributions of this work are summarized as follows.
\begin{itemize}
	
	\item A dual-BS rotatable array system assisted by a multi-layer active-passive transmitting RIS (T-RIS) is proposed. The discrete array rotations and positions are considered, and the SR maximization problem under sensing performance constraints is formulated. The framework aims to enhance secure transmission through joint optimization of the array pose, antenna distribution, beamforming matrices, and power allocation factor.
	
	\item To solve the non-convex optimization challenges, two algorithms are developed. First, a low-complexity online algorithm based on the generalized Rayleigh quotient (GRQ) is proposed, which optimizes the array pose and antenna distribution by joint beamforming design, followed by the sequential optimization of the sensing beam, power allocation factor, and T-RIS phase shift matrix.  To achieve higher security performance, an offline algorithm integrating the online scheme with the Multi-Agent Deep Deterministic Policy Gradient (MADDPG) framework is introduced, which enhances security performance through coordinated multi-agent learning.
	
	\item Simulation results demonstrate that the proposed scheme outperforms conventional systems without array pose adjustment and T-RIS assistance, achieving approximately 22\% improvement in SR. Furthermore, array rotation yields greater performance enhancement compared to positional adjustment.
\end{itemize}	
\subsection{Organization and Notation}

The following sections are organized as follows. Sec. \ref{s2} presents the system model. A low-complexity two-stage optimization scheme is proposed in Sec. \ref{s3}. Subsequently, Sect. \ref{s4} introduces a MADDPG-based optimization scheme. Simulation results are presented in Sec. \ref{s5}, and conclusion is provided in Sec. \ref{s6}.

Notations: Scalars, vectors, and matrices are denoted by lowercase (e.g., $t$), bold lowercase (e.g., $\mathbf{t}$), and bold uppercase (e.g., $\mathbf{T}$) letters, respectively. $|\cdot|$, $\|\cdot \|_0$, $\|\cdot \|$, and $\|\cdot \|_{F}$ denote the absolute value, $\ell_0$-norm, $\ell_2$-norm, and Frobenius norm, respectively. $[\cdot ]^{-1}$, $[\cdot ]^T$, and $[\cdot ]^H$ represent the matrix inverse, the transpose, and conjugate transpose, respectively. $[\cdot ]^+$ symbolizes the non-negative value. $[\cdot]_{n,m}$ denotes the $(n,m)$-th element of a matrix. $[\cdot]_{q}$ denotes the $q$-th element of a vector. $[\cdot]_{:,m}$ denotes the $m$-th column of a matrix. $\mathbb{E}\{\cdot \}$ stands for the expectation operator. The Hadamard product and Kronecker product are represented by $\odot$ and $\otimes$, respectively. $\text{diag}(\cdot )$ forms a diagonal matrix from its vector. $\text{conj}(\cdot )$ denotes the conjugate operator. The superscript $^*$ denotes the optimal value or solution. 
%$\angle$ denotes the phase in radians, and 
$\text{Tr}(\cdot)$ represents the trace of the matrix. 
\section{System Model}\label{s2}

\begin{figure}[t]
	\centering
	\includegraphics[width=0.4\textwidth, trim = 20 40 10 10,clip]{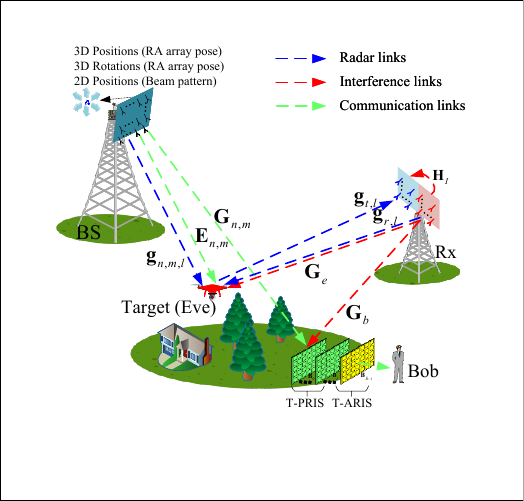}\\
	\caption{System model for proposed sensing-enabled secure Network with multi-layer T-RIS.}\label{f1}
\end{figure}
\begin{figure}[t]
	\centering
	\includegraphics[width=0.4\textwidth, trim = 10 20 10 20,clip]{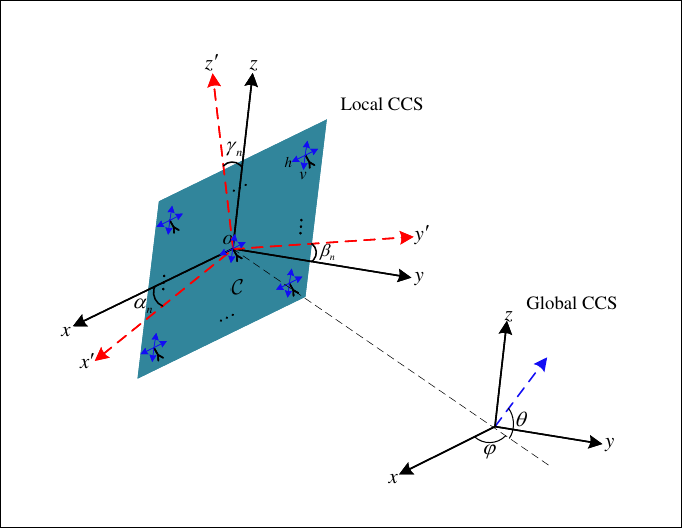}\\
	\caption{Illustration of position adjustment for the RA array and antennas.}\label{f2}
\end{figure}
	The system model is illustrated in Fig.~\ref{f1}. The system model comprises a BS, a sensing receiver, a target Eve, a multi-layer active-passive T-RIS, and a legitimate user (Bob). Specifically, the sensing receiver utilizes signals reflected from Eve to estimate its direction of arrival (DOA)  while transmitting AN with a wide beam pattern for simultaneous tracking and jamming. The BS transmits modulated communication signals to Bob while assisting the Rx, ensuring accurate estimation of Eve's DOA. Regarding antenna configuration, the BS employs a RA array controlled by a central processing unit, comprising $A$ antennas. Each antenna is adjustable in two-dimensional (2D) planar position, while the entire array is capable of three-dimensional (3D) positional and 3D rotational adjustments. Bob is equipped with $A_b$ FPAs. A total of $B$ RISs are deployed: $B-1$ passive RISs with $N_R$ elements are indexed by $\hat{b} \in \{0, 1,\ldots,B-2\}$, and one active RIS with $N_R$ elements is indexed by $\hat{b} = B-1$, where $\hat{b} \in \mathcal{B}=\{0, 1,\ldots,B-1\}$. Let $\hat{a}\in\{0, 1,\ldots,A-1\}$ denote the index set of MAs. The Rx is equipped with $A_r$ receiving and $A_t$ transmitting FPAs. Eve with $A_e$ FPAs is considered as a point target. Note that by leveraging orthogonal techniques, the proposed model can be extended to scenarios with multiple Bobs and multiple Eves. 

In contrast to 6D MA designs~\cite{Shao2025a}, the proposed RA array offers greater rotational degrees of freedom, thereby making it particularly tailored for secure communication in scenarios with sparsely distributed users or for dedicated service to specific users. Furthermore, the distribution of antenna elements on the array surface can be dynamically configured according to specific scenarios, enabling flexible beam pattern design. 
\subsection{Channel Model}
We consider discrete antenna positions, where each antenna element can only occupy limited locations on the given array plane, with both the array's position and rotation being adjusted at a given step size. As illustrated in Fig. \ref{f2}, we assume that there are $\hat{Q}$ discrete positions on the RA array (i.e., the 2D plane $\mathcal{C}$), denoted by set $\mathcal{\hat{Q}}= \{0,1,\ldots,\hat{Q}-1\}$, and RA array has $M$ discrete positions and $N$ discrete rotations, which are represented as sets $\mathcal{M}= \{0,1,\ldots,M-1\}$ and $\mathcal{N}= \{0,1,\ldots,N-1\}$, respectively. The set of candidate positions is represented as $\mathcal{P}= \{0,1,\ldots,P-1\}$, where $P=MN\hat{Q}$ denotes the total number of candidate positions.
%Let $\mathcal{C}=\{\mathbf{c}_0,\mathbf{c}_1,\ldots,\mathbf{c}_{Q-1}\}$ represents the set of all discrete positions on the array plane, where $\mathbf{c}_q$ ($q\in\mathcal{Q}$) represents the coordinate corresponding to the $q$-th discrete position. 

The set of all discrete positions on the array plane $\mathcal{C}$ can be expressed as 
\begin{equation}\label{1}
	\bm{\chi}=[\mathbf{t}_0,\mathbf{t}_1,\ldots,\mathbf{t}_{\hat{Q}-1}]^T, 
\end{equation}
where $\mathbf{t}_q = [x_q, 0, z_q]^T$ represents the coordinate corresponding to the $q$-th ($q\in\mathcal{\hat{Q}}$) discrete position of antenna. Note that antennas are initially distributed in the x-o-z plane (i.e., $y=0$). $x_q$ and $z_q$ represent the horizontal and vertical coordinates on $\mathcal{C}$, respectively.
%where $\mathbf{t}_q = [h_q, \nu_q]^T$ represents the coordinate corresponding to the $q$-th ($q\in\mathcal{\hat{Q}}$) discrete position of the MA element. $h_a$ and $\nu_a$ represent the horizontal and vertical coordinates on $\mathcal{C}$, respectively. %Consequently, \eqref{1} characterizes the spatial distribution of MA elements across  $\mathcal{C}$. 
%Without loss of generality, the MA elements can be initially distributed in the x-o-z plane (i.e., $y=0$). Thus, the position in the global Cartesian coordinate system (CCS) corresponding to the $q$-th discrete position of the MA element can be rewritten as $\mathbf{t}_q = [x_q, 0, z_q]^T$.

The $m$-th discrete position of the entire RA array can be represented by
\begin{equation}\label{2}
	\bm{\chi}_m=[\mathbf{t}_{m,0},\mathbf{t}_{m,1},\ldots,\mathbf{t}_{m,\hat{Q}-1}]^T, 
\end{equation}
where $\mathbf{t}_{m,q} = [x_{m,q}, y_{m,q}, z_{m,q}]^T$ $(m\in\mathcal{M})$. 
%Here, $x_{m,q}$, $y_{m,q}$ and $z_{m,q}$ denote the displacement components along the $x$-axis, $y$-axis, and $z$-axis, respectively, corresponding to the $q$-th candidate position of the $m$-th discrete position.
Then, the $q$-th discrete position of antenna after $n$-th ($n\in\mathcal{N}$) rotation can be expressed as~\cite{Shao2025}
\begin{align}
	{x}_{n,m,q}=&\ c_{\beta_n^{(m)}}c_{\gamma_n^{(m)}}x_q-s_{\beta_n^{(m)}}z_q,\\
	{y}_{n,m,q}=&\ (s_{\beta_n^{(m)}}s_{\alpha_n^{(m)}}c_{\gamma_n^{(m)}}-c_{\alpha_n^{(m)}}s_{\gamma_n^{(m)}})x_q\notag\\&+c_{\beta_n^{(m)}}s_{\alpha_n^{(m)}}z_q,\\
	{z}_{n,m,q}=&\ (c_{\alpha_n^{(m)}}s_{\beta_n^{(m)}}c_{\gamma_n^{(m)}}+s_{\alpha_n^{(m)}}s_{\gamma_n^{(m)}})x_q\notag\\&+(c_{\alpha_n^{(m)}}c_{\beta_n^{(m)}})z_q.
\end{align}
where $c_\imath=\cos(\imath)$ and $s_\imath=\sin(\imath)$. $\alpha_n$, $\beta_n$, and $\gamma_n$ are illustrated in Fig. \ref{f2}. Correspondingly, the candidate coordinates for antenna positions corresponding to the $m$-th position and the $n$-th rotation can be denoted as
\begin{equation}\label{6}
	\bm{\chi}_{n,m}(\bm{\chi})=[\mathbf{t}_{n,m,0}(\mathbf{t}_{0}),\mathbf{t}_{n,m,1}(\mathbf{t}_{1}),\!\ldots,\!\mathbf{t}_{n,m,\hat{Q}-1}(\mathbf{t}_{\hat{Q}-1})]^T, 
\end{equation}
where 
%\begin{equation}\label{7}
$\mathbf{t}_{n,m,q}(\mathbf{t}_{q}) = \mathbf{t}_{m,q}+[x_{n,m,q}, y_{n,m,q}, z_{n,m,q}]^T$
%\end{equation}
represents the coordinate corresponding to the $q$-th discrete position of antenna in the global Cartesian coordinate system. Then, the coordinate index set of all discrete positions and rotations corresponding to the RA array can be expressed as
%\begin{equation}
%	\begin{split}
%		\mathbf{T}=\frac{(\mathbf{T}_{\text{all}}\cdot\hat{\mathbf{T}})\otimes \mathbf{F}}{\|\hat{\mathbf{T}}\|_F},
%	\end{split}
%\end{equation}
\begin{equation}\label{8}
	\begin{split}
		\mathbf{T}_{\text{all}}=\mathbf{T}\otimes \mathbf{F},
	\end{split}
\end{equation}
where 
\begin{equation}
	\mathbf{T}=\begin{bmatrix}t_{0,0}&t_{0,1}&\cdots&t_{0,M-1}\\
		t_{1,0}&t_{1,1}&\cdots&t_{1,M-1}\\
		\vdots&\vdots&\ddots&\cdots\\
		t_{N-1,0}&t_{N-1,1}&\cdots&t_{N-1,M-1}\\\end{bmatrix}.
\end{equation}
Here, the coordinate index $t_{n,m}\in\{0,1\}$  can be mapped to the $n$-th rotation and $m$-th position of the RA array. 
%The matrix $\hat{\mathbf{T}}\in\mathbb{C}^{N\times M}$, composed of entirely distinct real elements, is employed to select discrete positions and rotations. 
%Therefore, the set of all discrete positions of $A$ MAs can be denoted as
%\begin{equation}
%	\begin{split}
%		\mathbf{T}=[\mathbf{T}_{0},\mathbf{T}_{1},\ldots,\mathbf{T}_{A-1}]^T\otimes \mathbf{F},
%	\end{split}
%\end{equation}
$\mathbf{F}=\text{diag}(\mathbf{f})\in\mathbb{C}^{\hat{Q}\times \hat{Q}}$ represents the position distribution of antennas on the array plane $\mathcal{C}$, satisfying $\|\mathbf{F}\|_0=A$, where $\mathbf{f}=[{f}_0,{f}_1,\ldots,{f}_{\hat{Q}-1}]^T $ $(f_q\in\{0,1\}, q\in\mathcal{\hat{Q}})$ and mapping one-to-one with $\bm{\chi}$. Given $\mathbf{F}$, all antennas change in the same position and rotation as the array performs. Thus, we have $\|\mathbf{T}\|_0=1$. For brevity, $\bm{\chi}_{n,m}\triangleq\bm{\chi}_{n,m}(\bm{\chi})$ and $\mathbf{t}_{n,m,q}\triangleq\mathbf{t}_{n,m,q}(\mathbf{t}_{q})$.
%Thus, we have $\|\mathbf{T}_a\|_0=1$ and $\|\mathbf{T}_0\|=\|\mathbf{T}_1\|=\ldots=\|\mathbf{T}_{A-1}\|$.

The channels between the BS and the T-RIS with $L_b$ paths, as well as between the BS and the Eve with $L_e$, are considered. The path set can be represented as $\mathcal{L}_b=\{0,1,\ldots,L_b-1\}$ and $\mathcal{L}_e=\{0,1,\ldots,L_e-1\}$, respectively. For the T-RIS, the azimuth and elevation angles corresponding to the $\iota$-th ($\iota\in\mathcal{L}_b$) path can be represented as $\phi_{n,m,\iota}$ and $\theta_{n,m,\iota}$, respectively. The steering vector from BS to the zeroth RIS over path $\iota$ can be represented as~\cite{Shao2025a} 
\begin{align}
	&\hat{\mathbf{G}}_{n,m}=\begin{bmatrix}\sqrt{\varsigma_{n,0}}\mathbf{a}_{n,m,0},\cdots,\sqrt{\varsigma_{n,L_b-1}}\mathbf{a}_{n,m,L_b-1}\end{bmatrix}^T,
\end{align}
where $\varsigma_{n,\iota}=10^{\frac{\kappa(\phi_{n,m,\iota},\theta_{n,m,\iota})}{10}}$, $\kappa(\phi_{n,m,\iota},\theta_{n,m,\iota})$ denotes the effective antenna gain corresponding to the $n$-th rotation, and
\begin{align}
	&\mathbf{a}_{n,m,\iota}=\begin{bmatrix}e^{j\frac{2\pi}{\lambda}\mathbf{v}_{n,m,\iota}^T\mathbf{t}_{n,m,0}},\cdots,e^{j\frac{2\pi}{\lambda}\mathbf{v}_{n,m,\iota}^T\mathbf{t}_{n,m,\hat{Q}-1}}\end{bmatrix}^T.
\end{align}
Here, the pointing vector is given by
\begin{align}
&\mathbf{v}_{n,m,\iota}=\notag\\&[\sin(\theta_{n,m,\iota})\cos(\phi_{n,m,\iota}),\sin(\theta_{n,m,\iota})\sin(\phi_{n,m,\iota}),\cos(\theta_{n,m,\iota})]^{T}.
\end{align}
Then, the channel from BS to the zeroth RIS corresponding to the $m$-th position and the $n$-th rotation can be denoted as
\begin{align}
	&{\mathbf{G}}_{n,m}={\mathbf{G}}_{n,m}^H\mathbf{H}_{R},
\end{align}
where $\mathbf{H}_{R}=[\mathbf{h}_{0},\mathbf{h}_{1},\ldots,\mathbf{h}_{N_R-1}]$ denotes the channel coefficient from BS to the zeroth RIS, $\mathbf{h}_{n_R}=[h_0,h_1,\ldots,h_{L_b-1}]^T$, and $n_R\in\mathcal{N}_R=\{0,1,\ldots,N_R-1\}$. Similarly, the channel $\mathbf{E}_{n,m}$ from BS to Eve corresponding to the $m$-th position and the $n$-th rotation can be obtained. The multi-path channels from the Rx to Bob and Eve are denoted as $\mathbf{G}_b$ and $\mathbf{G}_e$~\cite{Tang2025}, respectively. The channels from BS to Eve and from Eve to Rx are represented as $\mathbf{g}_{n,m,l}$ ($l\in\{0,2,\ldots,L_r-1\}$) and $\mathbf{g}_{r,l}$, respectively. $\mathbf{g}_{t,l}$ denotes the channel form Rx to Eve corresponding to $l$-th path.

The phase shift matrix of the $\hat{b}$-th ($\hat{b} \in \{0, 1,\ldots,B-2\}$) transmitting passive RIS (T-PRIS) can be expressed as $\bm{\Theta}_{\hat{b}}=\text{diag}(\bm{\theta}_{\hat{b}})$, where $\bm{\theta}_{\hat{b}}=[e^{j\theta_{0,{\hat{b}}}},e^{j\theta_{1,{\hat{b}}}},\ldots,e^{j\theta_{N_R-1,{\hat{b}}}}]^T$, $\theta_{n_R,\hat{b}}\in[0,2\pi)$. The phase shift matrix of the transmitting active RIS (T-ARIS) can be expressed as $\bm{\Theta}_{B-1}=\text{diag}(\bm{\theta}_{B-1})$, where $\bm{\theta}_{B-1}=[\beta_0 e^{j\theta_{0,B-1}},\beta_1 e^{j\theta_{1,B-1}},\ldots,\beta_{N_R-1} e^{j\theta_{N_R-1,B-1}}]^T$, $\theta_{\hat{n},B-1}\in[0,2\pi)$, and $\beta_{\hat{n}} \in[0,\beta_\text{max}]$. $\beta_\text{max}$ represents the maximum adjustable amplitude corresponding to the T-ARIS unit. For convenience, the channel from the zeroth RIS to Bob can be represented as \cite{Li2025}
\begin{equation}
	\mathbf{B}=\prod_{\hat{b}=0}^{B-1}\eta\bm{\Theta}_{\hat{b}}\mathbf{B}_{\hat{b}},
\end{equation}
where $\eta$ is the loss factor. $\mathbf{B}_{\hat{b}}$ denotes the channel between adjacent layers of the RIS.

\subsection{Communication Signal Model}
The modulated baseband signal for communication and sensing can be represented as
\begin{equation}
	\mathbf{x}=\mathbf{W}_c\mathbf{s}_{c}+\mathbf{W}_r\mathbf{s}_{r}=\sum_{i=0}^{\Upsilon_c-1}\mathbf{w}_is_i+\sum_{i=\Upsilon_c}^{\Upsilon_r+\Upsilon_c-1}\mathbf{w}_is_i=\mathbf{x}_c+\mathbf{x}_r,
\end{equation}
where $\mathbf{W}_c= [\mathbf{w}_0,\mathbf{w}_1,\ldots,\mathbf{w}_{\Upsilon_c-1}]\in\mathbb{C}^{\hat{Q}\times\Upsilon_c}$ and $\mathbf{W}_r=[\mathbf{w}_{\Upsilon_c},\mathbf{w}_{\Upsilon_c+1},\ldots,\mathbf{w}_{\Upsilon_r+\Upsilon_c-1}]\in\mathbb{C}^{\hat{Q}\times\Upsilon_r}$ are the beamforming matrices corresponding to communication and sensing, respectively. $\mathbf{s}_{c}= [{s}_0,{s}_1,\ldots,{s}_{\Upsilon_c-1}]^T\in\mathbb{C}^{\Upsilon_c\times1}$ and $\mathbf{s}_{r}[{s}_{\Upsilon_c},{s}_{\Upsilon_c+1},\ldots,{s}_{\Upsilon_r+\Upsilon_c-1}]^T\in\mathbb{C}^{\Upsilon_r\times1}$ denote communication and sensing signals, respectively, $\mathbb{E}\{\mathbf{s}_{c}\mathbf{s}_{c}^H\}=\mathbf{I}_{\Upsilon_c}$, $\mathbb{E}\{\mathbf{s}_{r}\mathbf{s}_{r}^H\}=\mathbf{I}_{\Upsilon_r}$. The AN signals transmitted at Rx can be represented by 
\begin{equation}
	\mathbf{x}_a=\mathbf{W}_a\mathbf{s}_{a},
\end{equation}
where $\mathbf{W}_a\in\mathbb{C}^{A_r\times\Upsilon_a}$ denotes the beamforming matrices corresponding to AN $\mathbf{s}_{a}\in\mathbb{C}^{\Upsilon_a\times 1}$. Here, $\Upsilon_c$, $\Upsilon_r$, and $\Upsilon_a$ denote the number of supported communication data streams at the BS, sensing data streams at the BS, and AN data streams at Rx, respectively. 

Then, the received signal $y_{b,n,m}$ corresponding to Bob can be expressed as
\begin{equation}\label{15}
	\begin{aligned}
	&y_{b,n,m}\\&=\sqrt{\alpha P_t}\mathbf{u}_b^H\mathbf{B}^H\mathbf{G}_{n,m}^H\mathbf{F}\mathbf{x}+\sqrt{(1-\alpha)P_t}\mathbf{u}_b^H\mathbf{B}^H\mathbf{G}_{b}^H\mathbf{x}_a\\&\quad+\mathbf{u}_b^H\mathbf{B}_{B-1}^H\mathbf{n}_I+\mathbf{u}_b^H\mathbf{n}_b\\
	&=\underbrace{\sqrt{\alpha P_t}\mathbf{u}_b^H\mathbf{B}^H\mathbf{G}_{n,m}^H\mathbf{F}\mathbf{W}_c\mathbf{s}_{c}}_{\text{\fontsize{8pt}{8pt}\selectfont Desired signal}}\\
	&\quad+\underbrace{\sqrt{\alpha P_t}\mathbf{u}_b^H\mathbf{B}^H\mathbf{G}_{n,m}^H\mathbf{F}\mathbf{W}_r\mathbf{s}_{r}\!+\!\!\sqrt{(\!1-\alpha\!)P_t}\mathbf{u}_b^H\mathbf{B}^H\mathbf{G}_{b}^H\mathbf{W}_a\mathbf{s}_{a}}_{\text{\fontsize{8pt}{8pt}\selectfont  Interference}}\\&\quad+\underbrace{\mathbf{u}_b^H\mathbf{B}_{B-1}^H\mathbf{n}_I+\mathbf{u}_b^H\mathbf{n}_b}_{\text{\fontsize{8pt}{8pt}\selectfont Noise}}.
	\end{aligned}
\end{equation}
where $P_t$, $0\leq\alpha\leq 1$, and $\mathbf{u}_b$ denote the transmission power threshold, power allocation factor, and receiving beamforming vector, respectively. $\mathbf{n}_b\sim\mathcal{CN}(\mathbf{0},\sigma_b^2\mathbf{I}_{A_b})$ symbolizes the additive white Gaussian noise (AWGN) for Bob. $\mathbf{n}_I\sim\mathcal{CN}(\mathbf{0},\sigma_I^2\mathbf{I}_{N_R})$ is the AWGN  thermal noise for the T-ARIS. Similarly, the received signal $y_{e,n,m}$ corresponding to Eve can be expressed as
\begin{equation}\label{16}
	\begin{aligned}
		y_{e,n,m}=&\ \sqrt{\alpha P_t}\mathbf{u}_e^H\mathbf{E}_{n,m}^H\mathbf{F}\mathbf{W}_c\mathbf{s}_{c}+\sqrt{\alpha P_t}\mathbf{u}_e^H\mathbf{E}_{n,m}^H\mathbf{F}\mathbf{W}_r\mathbf{s}_{r}\\&+\sqrt{(1-\alpha)P_t}\mathbf{u}_e^H\mathbf{G}_{e}^H\mathbf{W}_a\mathbf{s}_{a}+\mathbf{u}_e^H\mathbf{n}_e,
	\end{aligned}
\end{equation}
where $\mathbf{n}_e\sim\mathcal{CN}(\mathbf{0},\sigma_e^2\mathbf{I}_{A_e})$ symbolizes the AWGN for Eve. $\mathbf{u}_e$ represents the receiving beamforming vector corresponding to Eve, satisfying $\mathbf{u}_e^H\mathbf{u}_e=\mathbf{u}_b^H\mathbf{u}_b=1$. In MIMO systems where the number of antennas significantly exceeds that of users, it can be assumed that the different data streams are mutually orthogonal.
%In MIMO systems, multiple independent data streams can be transmitted to multiple users, with each user capable of receiving multiple streams. By employing techniques such as successive interference cancellation, the multi-stream signals can be effectively separated. For analytical tractability, we consider the scenario where Bob demodulates a single data stream.

With $y_{b,n,m}$ and $y_{e,n,m}$, the received signals $\mathbf{Y}_b\in\mathbb{C}^{N\times M}$ and $\mathbf{Y}_e\in\mathbb{C}^{N\times M}$ under all positions and rotations can be denoted as
	\begin{align}
	\mathbf{Y}_b
	&=\mathbf{T}\odot\mathbf{Y}_{b,c}+\mathbf{T}\odot\mathbf{Y}_{b,r}\notag\\
	&\quad+\mathbf{T}\otimes(\sqrt{(1-\alpha)P_t}\mathbf{u}_b^H\mathbf{B}^H\mathbf{G}_{b}^H\mathbf{W}_a\mathbf{s}_{a})\notag\\&\quad+\mathbf{T}\otimes({\mathbf{u}_b^H\mathbf{B}_{B-1}^H\mathbf{n}_I+\mathbf{u}_b^H\mathbf{n}_b}),\label{19}\\
	\mathbf{Y}_e&=\mathbf{T}\odot\mathbf{Y}_{e,c}+\mathbf{T}\odot\mathbf{Y}_{e,r}\notag\\&\quad+\mathbf{T}\otimes(\sqrt{(1-\alpha)P_t}\mathbf{u}_e^H\mathbf{G}_{e}\mathbf{W}_a\mathbf{s}_{a}+\mathbf{u}_e^H\mathbf{n}_e),\label{20}
\end{align}
where 
%$\mathbf{U}_b=\mathbf{1}_{N\times M}\otimes(\mathbf{u}_b^H\mathbf{B}^H)\in\mathbb{C}^{N\times MN_R}$, 
%$\mathbf{X}_c=\mathbf{1}_{N\times M}\otimes(\mathbf{W}_c\mathbf{s}_{c})\in\mathbb{C}^{N\hat{Q}\times M}$, 
\begin{align}
[\mathbf{Y}_{b,c}]_{n,m}&=\sqrt{\alpha P_t}\mathbf{u}_b^H\mathbf{B}^H\mathbf{G}_{n,m}^H\mathbf{F}\mathbf{W}_c\mathbf{s}_{c}, \\
[\mathbf{Y}_{b,r}]_{n,m}&=\sqrt{\alpha P_t}\mathbf{u}_b^H\mathbf{B}^H\mathbf{G}_{n,m}^H\mathbf{F}\mathbf{W}_r\mathbf{s}_{r}, \\
[\mathbf{Y}_{e,c}]_{n,m}&=\sqrt{\alpha P_t}\mathbf{u}_e^H\mathbf{E}_{n,m}^H\mathbf{F}\mathbf{W}_c\mathbf{s}_{c}, \\ [\mathbf{Y}_{e,r}]_{n,m}&=\sqrt{\alpha P_t}\mathbf{u}_e^H\mathbf{E}_{n,m}^H\mathbf{F}\mathbf{W}_r\mathbf{s}_{r}.
\end{align}
%$\mathbf{X}_r=\mathbf{1}_{N\times M}\otimes(\mathbf{W}_r\mathbf{s}_{r})\in\mathbb{C}^{N\hat{Q}\times M}$ are the block matrices. 
%$\mathbf{G}\in\mathbb{C}^{N\hat{Q}\times MN_R}$ and $\mathbf{E}\in\mathbb{C}^{N\hat{Q}\times MA_e}$ are block matrices, with their $(n, m)$-th elements given by $\mathbf{G}_{n,m}$ and $\mathbf{E}_{n,m}$, respectively.

%The SR is considered the performance metric for communication to ensure security.
 According to \eqref{19}, the SINR at Bob can be expressed as
\begin{equation}
	\text{SINR}_u=\frac{\|\mathbf{T}\odot\mathbf{Y}_{b,c}\|_{F}^2}{\|\mathbf{T}\odot\mathbf{Y}_{b,r}\|_{F}^2+\sum_{i=1}^{2}J_i},
\end{equation}
where 
\begin{align}
	J_1 &= (1-\alpha)P_t\|\mathbf{u}_b^H\mathbf{B}^H\mathbf{G}_{b}^H\mathbf{W}_a\|_F^2,\\
	J_2 &=\sigma_I^2\|\mathbf{u}_b^H\mathbf{B}_{B-1}^H\|_F^2+\sigma_b^2.
\end{align}	
According to \eqref{20}, the SINR at Eve can be expressed as
\begin{equation}
	\text{SINR}_e=\frac{\|\mathbf{T}\odot\mathbf{Y}_{e,c}\|_{F}^2}{\|\mathbf{T}\odot\mathbf{Y}_{e,r}\|_{F}^2+J_3+\sigma_e^2}
\end{equation}
where 
%\begin{align}
	$J_{3} = (1-\alpha)P_t\|\mathbf{u}_e^H\mathbf{G}_{e}^H\mathbf{W}_a\|_F^2$.
%\end{align}	
Then, the SR can be denoted as
\begin{align}
	R_s &= [\log_2(1+\text{SINR}_u)-\log_2(1+\text{SINR}_e)]^+\notag\\&=[r_b-r_e]^+.
\end{align}	
\subsection{Sensing Signal Model}
The echo signal reflected from the target and received at Rx can be represented by 
\begin{equation}
	\begin{aligned}
		&\mathbf{y}_{r,n,m}\!=\!\underbrace{\rho_0\mathbf{g}_{r,0}(\sqrt{\alpha P_t}\mathbf{g}_{n,m,0}^H\mathbf{F}\mathbf{W}_r\mathbf{s}_r+\sqrt{(1-\alpha) P_t}\mathbf{g}_{t,0}^H\mathbf{W}_a\mathbf{s}_a)}_{\text{\fontsize{8pt}{8pt}\selectfont Desired echo signal}}\\&+\underbrace{\sum_{l=1}^{L_r-1}\rho_l\mathbf{g}_{r,l}(\sqrt{\alpha P_t}\mathbf{g}_{n,m,l}^H\mathbf{F}\mathbf{W}_r\mathbf{s}_r+\sqrt{(1-\alpha) P_t}\mathbf{g}_{t,l}^H\mathbf{W}_a\mathbf{s}_a)}_{\text{\fontsize{8pt}{8pt}\selectfont Clutter}}\\&+\underbrace{g_I\mathbf{H}_I\mathbf{x}_a+\sum_{l=0}^{L_r-1}\sqrt{\alpha P_t}\rho_0\mathbf{g}_{r,l}\mathbf{g}_{n,m,l}^H\mathbf{F}\mathbf{W}_c\mathbf{s}_c}_{\text{\fontsize{8pt}{8pt}\selectfont Interference}}+\underbrace{\mathbf{n}_r}_{\text{\fontsize{8pt}{8pt}\selectfont Noise}},
	\end{aligned}
\end{equation}
where $\rho_l$ denotes the reflection coefficient corresponding to the $l$-th path. $\mathbf{H}_I$ denotes the loop interference channel. $g_I$ represents theresidual self interference coefficient.  %$\mathbf{u}_r$ represents the receiving beamforming vector corresponding to Rx, satisfying $\mathbf{u}_r^H\mathbf{u}_r=1$.
 $\mathbf{n}_r\sim\mathcal{CN}(\mathbf{0},\sigma_r^2\mathbf{I}_{A_r})$ symbolizes the AWGN for Rx.

%For $M$ positions and $N$ rotations,  $\mathbf{G}_l=[\mathbf{g}_{n,m,l}]_{\forall n,m}\in\mathbb{C}^{N\hat{Q}\times M}$ is constructed to represent the $MN$ possible channels from BS to the target. 
With $y_{r,n,m}$, the received signals $\mathbf{Y}_r$ under all positions and rotations can be denoted as
%\begin{equation}
	\begin{align}
		\mathbf{Y}_r&=\mathbf{T}\odot\mathbf{Y}_{r,1}+\mathbf{T}\otimes(\rho_0\sqrt{(1-\alpha) P_t}\mathbf{g}_{r,0}\mathbf{g}_{t,0}^H\mathbf{W}_a\mathbf{s}_a)\notag\\&+\mathbf{T}\odot\mathbf{Y}_{r,2}+\mathbf{T}\otimes(\sum_{l=1}^{L_r-1}\rho_l\sqrt{(1-\alpha) P_t}\mathbf{g}_{r,l}\mathbf{g}_{t,l}^H\mathbf{W}_a\mathbf{s}_a)\notag\\&+\mathbf{T}\otimes({g_I\mathbf{H}_I\mathbf{x}_a}+\sum_{l=0}^{L_r-1}\rho_0\sqrt{\alpha P_t}\mathbf{g}_{r,l}\mathbf{g}_{n,m,l}^H\mathbf{F}\mathbf{W}_c\mathbf{s}_c+{\mathbf{n}_r}),\label{32}
	\end{align}
%\end{equation}
where 
\begin{align}
[\mathbf{Y}_{r,1}]_{nA_r:(n+1)A_r-1,m}&=\rho_0\sqrt{\alpha P_t}\mathbf{g}_{r,0}\mathbf{g}_{n,m,0}^H\mathbf{F}\mathbf{W}_r\mathbf{s}_r,\\
[\mathbf{Y}_{r,2}]_{nA_r:(n+1)A_r-1,m}&=\sum_{l=1}^{L_r-1}\rho_l\sqrt{\alpha P_t}\mathbf{g}_{r,l}\mathbf{g}_{n,m,l}^H\mathbf{F}\mathbf{W}_r\mathbf{s}_r.
\end{align}
%$\mathbf{U}_{r,l}=\mathbf{1}_{N\times M}\otimes\mathbf{g}_{r,l}\in\mathbb{C}^{NA_r\times M}$ and $\hat{\mathbf{X}}_{a,l}=\mathbf{1}_{M\times M}\otimes(\mathbf{g}_{t,l}^H\mathbf{x}_a)\in\mathbb{C}^{M\times M}$ are the constructed block matrices. 
%For clarity, the main notations are listed in Table \ref{tab}.

The root minimum mean square error (RMMSE) of DOA estimation can be denoted as~\cite{Liu2025}
\begin{equation}\label{33}
	\epsilon_{\theta}=\frac{\theta_{\text{3dB}}}{1.6\sqrt{2\mathrm{SINR_r}}}, \epsilon_{\phi}=\frac{\phi_{\text{3dB}}}{1.6\sqrt{2\mathrm{SINR_r}}},
\end{equation}
where $\theta_{\text{3dB}}$ and $\phi_{\text{3dB}}$ denote the 3dB beamwidths, respectively. According to \eqref{32}, the SINR at Rx can be expressed as
%\begin{equation}
	$\text{SINR}_r = \frac{J_4}{\sum_{i=5}^{7}J_i}$,
%\end{equation}
where 
\begin{align}
	J_4 &= \|\mathbf{T}\odot\mathbf{Y}_{r,1}\|_F^2+\|\mathbf{T}\otimes(\rho_0\sqrt{(1-\alpha) P_t}\mathbf{g}_{r,0}\mathbf{g}_{t,0}^H\mathbf{W}_a)\|_{F}^2,\\
	J_5 &= \|\mathbf{T}\odot\mathbf{Y}_{r,2}\|_F^2+\|\mathbf{T}\otimes(\!\sum_{l=1}^{L_r-1}\!\rho_l\!\sqrt{(1-\alpha) P_t}\mathbf{g}_{r,l}\mathbf{g}_{t,l}^H\mathbf{W}_a\!)\|_F^2,\\
	J_{6} &= \|\mathbf{T}\otimes\sum_{l=0}^{L_r-1}\rho_0\sqrt{\alpha P_t}\mathbf{g}_{r,l}\mathbf{g}_{n,m,l}^H\mathbf{F}\mathbf{W}_c\|_F^2+A_r\sigma_r^2,\\
	J_{7} &=g_I\|{\mathbf{H}_I\mathbf{x}_a}\|_F^2.
\end{align}	

\subsection{Problem Formulation}
We aim to maximize the SR under the constraint of highly accurate angle sensing. Let $\Gamma_{\theta}$ and $\Gamma_{\phi}$ symbolize the thresholds for the RMMSE of the corresponding DOA estimations. The angle estimation requirements are expressed as $\epsilon_{\theta}\leq \Gamma_{\theta}$ and $\epsilon_{\phi}\leq \Gamma_{\phi}$. Based on \eqref{33}, we have
\begin{align}
	\text{SINR}_r\geq\Gamma_r,\Gamma_r = \max\{\frac{\theta_{\text{3dB}}^2}{1.6^2\Gamma_{\theta}^2},\frac{\phi_{\text{3dB}}^2}{1.6^2\Gamma_{\phi}^2}\}.
\end{align}
%We consider a target tracking scenario where the BS and Rx jointly predict the target's azimuth and elevation angles in the first stage. In the second stage, Rx performs hybrid active-passive DOA estimation to achieve precise target localization. 
Let $\mathbf{W}\in\{\mathbf{W}_c,\mathbf{W}_r,\mathbf{W}_a\}$ and $\mathbf{u}\in\{\mathbf{u}_b,\mathbf{u}_e\}$ represent the sets of the precoding matrices and the received beamforming vectors, respectively. Consequently, the optimization problem can be formulated as
\begin{subequations}
	\begin{align}
		\label{eq34a}
		\text{P1:}\quad\quad&\underset{\mathbf{W},\alpha,\mathbf{B},\mathbf{T},\mathbf{F},\mathbf{u}}{\max}\ R_s
		\\ \label{eq34b}
		\mathrm{s.t.}\  &\text{C1}:\text{Tr}(\mathbf{W}_{c}^H\mathbf{F}\mathbf{W}_{c})+\text{Tr}(\mathbf{W}_{r}^H\mathbf{F}\mathbf{W}_{r}) = 1, \\ 
		&\text{C2}:\text{Tr}(\mathbf{W}_{a}^H\mathbf{F}\mathbf{W}_{a}) = 1, \\ 
		%		&\quad\ \hat{\mathbf{\hat{Q}}}_k\mathbf{v}_k=\mathbf{0}, \forall k\\
		&\text{C3}:\mathbf{u}_b^H\mathbf{u}_b=1,\\&\text{C4}:\mathbf{u}_{e}^H\mathbf{u}_{e}=1,\\
		%&\text{C4}:\epsilon_{\theta}\leq \Gamma_{\theta},\epsilon_{\phi}\leq \Gamma_{\phi},\\
		&\text{C5}:\text{SINR}_r\geq \Gamma_r,\\
		&\text{C6}:\mathbf{u}_b^H\mathbf{B}^H\mathbf{G}_{b}^H\mathbf{W}_a\mathbf{s}_{a}=0,\\
		&\text{C7}:\|\mathbf{F}\|_0=A, \|\mathbf{T}\|_0=1,\notag\\&\qquad t_{n,m}\in\{0,1\},f_q\in\{0,1\}, \forall n,m,q,\\\label{34h}
		&\text{C8}:\|\mathbf{t}_{n.m,\hat{a}}-\mathbf{t}_{n,m,\hat{a}^\prime}\|\geq d, \hat{a}\neq \hat{a}^\prime,\forall n,m,\\
		&\text{C9}:\|\sqrt{\alpha P_t}\bm{\Theta}_{{B}-1}^H(\prod_{\hat{b}=0}^{B-2}\eta\bm{\Theta}_{\hat{b}}\mathbf{B}_{\hat{b}})^H\mathbf{G}_{n,m}^H\mathbf{F}\mathbf{x}\|^2\notag\\&\qquad+\sigma_r^2\|\bm{\Theta}_{{B}-1}\|_F^2\leq P_{\text{RIS}},\forall n,m,
	\end{align}
\end{subequations}
where $P_{\text{RIS}}$ symbolizes the maximum power corresponding to T-RIS. C1 and C2 denote the transmit power constraints. C3 and C4 represent the receive beamforming constraint. C7 symbolizes the array pose adjustment and antenna distribution. C8 indicates that the minimum spacing of the antennas is $d$. Note that the rotation and position constraints of the RA array can be preconfigured to prevent signal blockage. 
%In the scenario where the BS obtains the elevation angle $\theta_l$ and depression angle B of the target and scatterers through channel estimation, multipath effects may introduce errors in the estimation of target number and angles at the receiver when using the MUSIC algorithm. Under a 3D space, the MUSIC spectrum can be expressed as
%\begin{equation}
%	p(\theta,\varphi)=\frac{1}{\mathbf{a}^{H}(\theta,\varphi)\mathbf{N}_s\mathbf{N}_s^{H}\mathbf{a}(\theta,\varphi)},
%\end{equation}
%where $\mathbf{N}_s$ represents the signal subspace. 

\section{Proposed Low-Complexity Two-Stage Optimization Scheme}\label{s3}

In this section, P1 is addressed via a two-stage method. The optimization problem is solved by first decomposing it into maximizing Bob's rate, then jointly maximizing the SR through power allocation and sensing optimization.
\subsection{Maximization of the Achievable Rate for Bob}
Specifically, the subproblem for the maximization of Bob's transmission rate can be formulated as
\begin{subequations}
	\begin{align}
		\label{eq41a}
		\text{P2:}\quad&\underset{\mathbf{W}_c,\mathbf{B},\mathbf{T},\mathbf{F},\mathbf{u}_b}{\max}\ \frac{1}{J_2}\|\mathbf{T}\odot\mathbf{Y}_{b,c}\|_{F}^2
		\\ 
		&\quad\mathrm{s.t.}\  \text{C1}^{\prime}:\text{Tr}(\mathbf{W}_{c}^H\mathbf{F}\mathbf{W}_{c}) \leq 1,\\&\quad\quad\ \ \text{C3},\text{C7},\text{C8},\text{C9}.
	\end{align}
\end{subequations}
Note that the sensing performance is optimized in the subsequent step. Therefore, $\|\mathbf{Y}_{b,r}\|_{F}^2$ can be regarded as a constant. Additionally, constraint C6 implies that $J_1$ can be omitted.

To solve P2, an alternating optimization method is employed. The initial value of $\mathbf{F}$ is set to an identity matrix to symbolize the potential full-array gain. The underlying concept is that, once the optimal sparse matrix $\mathbf{F}$ is obtained, a lower-dimensional identity matrix can be derived by selecting its non-zero values. $\mathbf{B}=\eta\mathbf{1}_{N_R\times A_b}$ signifies a scattering cluster, and $\alpha=1$ symbolizes the single-BS system configuration.
\subsubsection{Optimization of $\mathbf{T}$}
With $\mathbf{B}$, $\mathbf{F}$, and $\mathbf{u}_b$ fixed, the subproblem with respect to (w.r.t.) $\mathbf{W}_c$ and $\mathbf{T}$ is given by
\begin{subequations}
	\begin{align}
		\text{P3:}\quad&\underset{\mathbf{W}_c,\mathbf{T}}{\max}\ \|\mathbf{T}\odot\mathbf{Y}_{b,c}\|_{F}^2\label{eq44a}
		\\ 
		&\mathrm{s.t.}\  \text{C1}^{\prime},\\&\ \ \quad\text{C7}^{\prime}: \|\mathbf{T}\|_0=1,t_{n,m}\in\{0,1\},\forall n,m.
	\end{align}
\end{subequations}
By utilizing the constraint \( |\mathbf{T}|_0 = 1 \), \eqref{eq44a} can be rewritten as
\begin{align}
&\ \underset{\mathbf{W}_c,\mathbf{T}}{\max}\ \sum_{m=0}^{M-1}\sum_{n=0}^{N-1}\sum_{i=0}^{\Upsilon_c-1}t_{n,m}\mathbf{w}_i^H\mathbf{F}\mathbf{G}_{n,m}\mathbf{B}\mathbf{U}_b\mathbf{B}^H\mathbf{G}_{n,m}^H\mathbf{F}\mathbf{w}_i\notag\\
	\triangleq&\ \underset{\mathbf{W}_c,\mathbf{T}}{\max}\ \sum_{i=0}^{\Upsilon_c-1}\mathbf{w}_i^H \!\left( \sum_{m=0}^{M-1}\sum_{n=0}^{N-1}t_{n,m}\mathbf{F}\mathbf{G}_{n,m}\mathbf{B}\mathbf{U}_b\mathbf{B}^H\mathbf{G}_{n,m}^H\mathbf{F}\right)\!\mathbf{w}_i\notag\\
	\triangleq&\ \underset{\mathbf{W}_c,\mathbf{T}}{\max}\ \sum_{i=0}^{\Upsilon_c-1}\mathbf{w}_i^H\bm{{\varPsi}}_1\mathbf{w}_i,
\end{align}
where  
\begin{align}\label{46}
\bm{{\varPsi}}_1= \sum_{m=0}^{M-1}\sum_{n=0}^{N-1}t_{n,m}\mathbf{F}\mathbf{G}_{n,m}\mathbf{B}\mathbf{U}_b\mathbf{B}^H\mathbf{G}_{n,m}^H\mathbf{F}= \sum_{m=0}^{M-1}\sum_{n=0}^{N-1}\bm{{\varPsi}}_{n,m}
\end{align}
 and $\mathbf{U}_b=\mathbf{u}_b\mathbf{u}_b^H$. 
 
 To solve P3, two approaches are developed based on search and merge, respectively. The optimal position and rotation of the RA array are obtained by evaluating the objective function across all possible configurations. Specifically, P3 can be decomposed into a set of subproblems, each corresponding to the $n$-th rotation and the $m$-th position, formulated as
 \begin{subequations}\label{eq47}
 	\begin{align}
 		\text{P3.1:}\quad&\underset{\mathbf{w}_i}{\max}\ f_{n,m}=\mathbf{w}_i^H\bm{{\varPsi}}_{n,m}\mathbf{w}_i
 		\\ 
 		&\quad\mathrm{s.t.}\  \text{C1}^{\prime\prime}: \mathbf{w}_i^H\mathbf{F}\mathbf{w}_i\leq\frac{1}{\Upsilon_c}. 
 	\end{align}
 \end{subequations}
 Here, $i\in\{0,1,\ldots,\Upsilon_c-1\}$ denotes the index of the data stream. P3.1 can be solved via the GRQ. The optimal solution for $\mathbf{w}_i$ is given by the eigenvector corresponding to the largest eigenvalue of $\mathbf{F}^{-1}\bm{{\varPsi}}_{n,m}$. Then, the maximum value $f_{n,m}^{\max}$ of $f_{n,m}$ can be obtained. For all possible positions and rotations, $N×M$ maximum values can be obtained, i.e., $\{f_{n,m}^{\max}\}_{\forall n,m}$. The global maximum within $\{f_{n,m}^{\max}\}_{\forall n,m}$ corresponds to the optimal rotation and position.
 
 According to \eqref{46}, by aggregating the cascaded channels corresponding to all rotations and positions, P3 can be expressed as
\begin{subequations}\label{eq48}
	\begin{align}
		\text{P3.2:}\quad&\underset{\mathbf{w}_i,\mathbf{T}}{\max}\ \mathbf{w}_i^H\bm{{\varPsi}}_{1}\mathbf{w}_i\label{eq46a}
		\\ 
		&\quad\mathrm{s.t.}\  \text{C1}^{\prime\prime}, \text{C7}^{\prime}.
	\end{align}
\end{subequations}
By relaxing constraint C7$^\prime$, P3.2 can be solved via the GRQ. The optimal solution for $\mathbf{w}_i$ is given by the eigenvector corresponding to the largest eigenvalue of $\Upsilon_c\bm{{\varPsi}}_{1}^{-1}\mathbf{F}$. Subsequently, the value of $\mathbf{w}_i^H\bm{{\varPsi}}_{n,m}\mathbf{w}_i$ is computed for all possible rotations and positions. The optimal RA array configuration is thereby obtained by comparing these values. 
%The core idea of this approach is to select the strongest paths from the virtual aggregated channel $\bm{{\varPsi}}_{1}$ for transmission, whereby the configuration corresponding to the value of $\mathbf{w}_i^H\bm{{\varPsi}}_{n,m}\mathbf{w}_i$ that is closest to the objective function is chosen as the optimal rotation and position.

%Given $\{\mathbf{w}_i\}_{i=0}^{\Upsilon_c-1}$ and the optimal position and rotation, $\mathbf{W}_c$ and $\mathbf{T}$ can be obtained.
\subsubsection{Optimization of $\mathbf{W}_c$ and $\mathbf{F}$}
With $\mathbf{T}$, $\mathbf{B}$, and $\mathbf{u}_b$ fixed, the subproblem w.r.t. $\mathbf{F}$ and $\mathbf{W}_c$ can be given by
\begin{subequations}\label{49}
	\begin{align}
		\text{P4:}\quad&\underset{\mathbf{w}_i,\mathbf{F}}{\max}\ t_{n,m}^*\mathbf{w}_i^H\mathbf{F}\mathbf{G}_{n,m}^*\mathbf{B}\mathbf{U}_b\mathbf{B}^H(\mathbf{G}_{n,m}^*)^H\mathbf{F}\mathbf{w}_i\label{eq49a}
		\\ 
		&\mathrm{s.t.}\  \text{C1}^{\prime\prime},\text{C8},\\&\quad\ \ \text{C7}^{\prime\prime}: \|\mathbf{F}\|_0=A,f_{q}\in\{0,1\},\forall q,
	\end{align}
\end{subequations}
where $t_{n,m}^*=1$ and $\mathbf{G}_{n,m}^*$ denote the index and channel corresponding to the optimal $n$-th rotation and $m$-th position, respectively. Given that $\mathbf{G}_{n,m}^*$ collects all virtual and real channels compliant with the minimum antenna spacing, constraint C8 is redundant and can be disregarded.

While the optimization of position and rotation for a single RA array has polynomial complexity, the selection of $A$ optimal antenna positions from $\hat{Q}$ candidates is of exponential complexity, thus making an exhaustive search infeasible. To circumvent the high complexity, maximum ratio transmission (MRT) is adopted. Specifically, with $\mathbf{F}=\mathbf{I}_{\hat{Q}}$ fixed, $\mathbf{w}_i$ can be given by
\begin{align}\label{50}
\mathbf{w}_i =\frac{\mathbf{G}_{n,m}^*\mathbf{B}\mathbf{u}_b}{\sqrt{\Upsilon_c}\|\mathbf{G}_{n,m}^*\mathbf{B}\mathbf{u}_b\|}.
\end{align}
Substituting \eqref{50} into \eqref{49}, we obtain
\begin{subequations}\label{51}
	\begin{align}
		\text{P5:}\quad&\underset{\mathbf{f}}{\max}\ |\sum_{q=0}^{\hat{Q}-1}h_qf_q|^2
		\\ 
		&\quad\mathrm{s.t.}\  \text{C7}^{\prime\prime\prime}:\sum_{q=0}^{\hat{Q}-1}f_q=A,f_{q}\in\{0,1\},\forall q,
	\end{align}
\end{subequations}
where $\mathbf{h}_{n,m}^*=\mathbf{u}_b\mathbf{B}^H(\mathbf{G}_{n,m}^*)^H\text{diag}(\mathbf{G}_{n,m}^*\mathbf{B}\mathbf{u}_b)$ and $[\mathbf{h}_{n,m}^*]_q=h_q$. P5 can be solved via phase alignment (PA). Then, the top $A$ positions with the strongest channel gains are chosen as the antenna positions. With the optimal position selection matrix $\mathbf{F}^*=\text{diag}(\mathbf{f^*})$ for antennas, $\mathbf{w}_i$ can be normalized to
\begin{align}\label{52}
	\mathbf{w}_i^* =\frac{\mathbf{F}^{*}\mathbf{G}_{n,m}^*\mathbf{B}\mathbf{u}_b}{\sqrt{\Upsilon_c}\|\mathbf{F}^*\mathbf{G}_{n,m}^*\mathbf{B}\mathbf{u}_b\|}.
\end{align}
Given $\{\mathbf{w}_i^*\}_{i=0}^{\Upsilon_c-1}$, the beamforming matrix $\mathbf{W}_c^*$ can be obtained.
\subsubsection{Optimization of $\mathbf{B}$}
Similarly, with other variables fixed, the subproblem w.r.t. $\mathbf{B}$ is give by
\begin{subequations}\label{53}
	\begin{align}
		\text{P6:}\quad&\underset{\mathbf{B}}{\max}\ t_{n,m}^*(\mathbf{w}_i^*)^H\mathbf{F}^*\mathbf{G}_{n,m}^*\mathbf{B}\mathbf{U}_b\mathbf{B}^H(\mathbf{G}_{n,m}^*)^H\mathbf{F}^*\mathbf{w}_i^*
		\\ 
		&\mathrm{s.t.}\  \text{C9}.
	\end{align}
\end{subequations}
T-RIS is employed to aggregate and amplify the received power, thereby enhancing the quality of service for Bob. However, the unit-modulus constraint of the PRIS typically leads to high algorithmic complexity. To reduce complexity, the signal power is concentrated at the center of the next RIS by $B-1$ T-PRISs~\cite{Li2025}. Subsequently, interference is suppressed at the T-ARIS via leakage theory (LT). Specifically, the corresponding phase shift matrix $\bm{\Theta}_{0}$ for the zero-th T-PRIS can be given by
\begin{align}\label{eq54}
\bm{\Theta}_{0}^*=\text{diag}(\bm{\theta}_{0}^*)=\text{diag}(e^{j\angle(\text{diag}(\mathbf{G}_{n,m}^H\mathbf{F}\mathbf{x}_c))^{^{-1}}[\mathbf{B}_{0}]_{:,\zeta}}),
\end{align}
where $\zeta$ symbolizes the center of the next RIS. For the $\hat{b}$-th ($0\leq\hat{b}\leq B-2$) T-PRIS, the phase shift matrix $\bm{\Theta}_{\hat{b}}$ can be represented as
\begin{align}\label{eq55}
	\bm{\Theta}_{\hat{b}}^*&=\text{diag}(\bm{\theta}_{\hat{b}}^*)=\text{diag}(e^{\angle((\prod_{i=1}^{\hat{b}-1}\bm{\Theta}_{i}\mathbf{B}_{i})^H\mathbf{G}_{n,m}^H\mathbf{F}\mathbf{x}_c)^{-1}[\mathbf{B}_{\hat{b}}]_{:,\zeta}}).
\end{align}
%where 
%\begin{align}
%	\bm{\vartheta}_{\hat{b}}^*&=\text{diag}\bigg( \mathbf{u}_b^H \bm{\Theta}_{\hat{B}-1}^H(\prod_{i=\hat{b}+1}^{B-2}\eta\bm{\Theta}_{i}\mathbf{B}_{i})^H\bigg)\mathbf{B}_{\hat{b}}^H (\prod_{i=1}^{\hat{b}-1}\bm{\Theta}_{i}\mathbf{B}_{i})^H\notag\\&\quad\mathbf{B}_{0}^H\bm{\Theta}_{0}^H\mathbf{G}_{n,m}^H\mathbf{F}\mathbf{x}_c.
%\end{align}
Then, based on the LT, the optimization scheme for the phase shift matrix $\bm{\Theta}_{B-1}$ of the T-ARIS can be formulated as
\begin{subequations}\label{57}
	\begin{align}
		\text{P7:}\quad&\underset{\bm{\Theta}_{B-1}}{\max}\ \frac{\bm{\theta}_{B-1}^H\mathbf{C}\bm{\theta}_{B-1}}{\bm{\theta}_{B-1}^H\mathbf{D}\bm{\theta}_{B-1}}
		\\ 
		&\mathrm{s.t.}\  \text{C10}:\bm{\theta}_{B-1}^H\bm{\theta}_{B-1}=1,
	\end{align}
\end{subequations}
where
\begin{align}
	&\mathbf{C}=\text{conj}(\mathbf{H}_B\mathbf{B}_{B-1}\mathbf{U}_b\mathbf{B}_{B-1}^H\mathbf{H}_B^H),\\
	&\mathbf{D}=\text{conj}(\hat{\mathbf{H}}_B\hat{\mathbf{H}}_B^H+\frac{\sigma_r^2}{\alpha P_t}\mathbf{I}_{N_R}),\\
	&\mathbf{H}_B= \text{diag}\left((\mathbf{w}_i^*)^H\mathbf{F}^*\mathbf{G}_{n,m}^*\hat{\mathbf{B}}\right),
	\\ 
		&\hat{\mathbf{H}}_B= \text{diag}\left(\mathbf{x}^H\mathbf{F}^*\mathbf{G}_{n,m}^*\hat{\mathbf{B}}\right),\\
	&\hat{\mathbf{B}}=\prod_{\hat{b}=0}^{B-2}\eta\bm{\Theta}_{\hat{b}}\mathbf{B}_{\hat{b}}.
\end{align}
The closed-form solution to P7 is the eigenvector corresponding to the largest eigenvalue of $\mathbf{D}^{-1}\mathbf{C}$. Then, by collecting all the phase shift matrices, the channel matrix $\mathbf{B}^*$ can be obtained.

\subsubsection{Optimization of $\mathbf{u}_b$}
The MMSE criterion is employed for receive beamforming design, which yields
\begin{align}
\mathbf{U}_i^{\text{MSE}}&= \mathbb{E}\{(y_{b,n,m}-s_i)(y_{b,n,m}-s_i)^H\}\notag\\
&=\sum_{i=0}^{\Upsilon_r+\Upsilon_c-1}\sqrt{\alpha P_t}\mathbf{u}_b^H\mathbf{B}^H\mathbf{G}_{n,m}^H\mathbf{F}\mathbf{w}_i\mathbf{w}_i^H\mathbf{F}^H\mathbf{G}_{n,m}\mathbf{B}\mathbf{u}_b\notag\\&\quad+\sqrt{(1-\alpha)P_t}\mathbf{u}_b^H\mathbf{B}^H\mathbf{G}_{b}^H\mathbf{W}_a\mathbf{W}_a^H\mathbf{G}_{b}\mathbf{B}\mathbf{u}_b\notag\\&\quad-\sqrt{\alpha P_t}\mathbf{u}_b^H\mathbf{B}^H\mathbf{G}_{n,m}^H\mathbf{F}\mathbf{w}_i-\sqrt{\alpha P_t}\mathbf{w}_i^H\mathbf{F}^H\mathbf{G}_{n,m}\mathbf{B}\mathbf{u}_b\notag\\&\quad+\sigma_I^2\mathbf{u}_b^H\mathbf{B}_{B-1}^H\mathbf{B}_{B-1}\mathbf{u}_b+\sigma_I^2\mathbf{u}_b^H\mathbf{u}_b+1\notag\\&=\mathbf{u}_b^H\mathbf{U}_1\mathbf{u}_b-\sqrt{\alpha P_t}\mathbf{u}_b^H\mathbf{B}^H\mathbf{G}_{n,m}^H\mathbf{F}\mathbf{w}_i+U_2,
\end{align}
where
\begin{align}
	\mathbf{U}_1
	&=\sum_{i=0}^{\Upsilon_r+\Upsilon_c-1}\sqrt{\alpha P_t}\mathbf{B}^H\mathbf{G}_{n,m}^H\mathbf{F}\mathbf{w}_i\mathbf{w}_i^H\mathbf{F}^H\mathbf{G}_{n,m}\mathbf{B}\notag\\&\quad+\sqrt{(1-\alpha)P_t}\mathbf{B}^H\mathbf{G}_{b}^H\mathbf{W}_a\mathbf{W}_a^H\mathbf{G}_{b}\mathbf{B}\notag\\&\quad+\sigma_I^2\mathbf{B}_{B-1}^H\mathbf{B}_{B-1}+\sigma_I^2\mathbf{I}_{A_b},\\U_2&=-\sqrt{\alpha P_t}\mathbf{w}_i^H\mathbf{F}^H\mathbf{G}_{n,m}\mathbf{B}\mathbf{u}_b+1.
\end{align}
By taking the derivative of the covariance matrix $\mathbf{U}_i^{\text{MSE}}$ w.r.t. $\mathbf{u}_b^H$, we obtain
\begin{align}
	\mathbf{u}_b=\sqrt{\alpha P_t}\mathbf{U}_1^{-1}\mathbf{B}^H\mathbf{G}_{n,m}^H\mathbf{F}\mathbf{w}_i.
\end{align}	
Accordingly, $\mathbf{W}_c$, $\mathbf{B}$, $\mathbf{T}$, $\mathbf{F}$, and $\mathbf{u}_b$ are obtained, and the transmission rate corresponding to Bob can be calculated. The method for maximizing Bob's achievable rate is illustrated in Algorithm \ref{alg1}.
 \begin{algorithm}[t]
	\renewcommand{\algorithmicrequire}{\textbf{Input:}}
	\renewcommand{\algorithmicensure}{\textbf{Output:}}
	\caption{Two-stage Optimization Algorithm.}\label{alg1}
	\begin{algorithmic}[1]
		\REQUIRE $P_t$, $\Gamma_r$, $d$, $P_{\text{RIS}}$, $\Upsilon_c$, $\Upsilon_r$, $\Upsilon_a$, $\theta_{\text{3dB}}$, $\phi_{\text{3dB}}$, $\sigma_b$, $\sigma_e$, $\sigma_r$, $\{\varsigma_{n,\iota}\}_{\iota=0}^{L_b-1}$, $\{\rho_l\}_{l=0}^{L_r-1}$, $\{\mathbf{g}_{n,m,l}\}_{l=0}^{L_r-1}$, $\{\mathbf{g}_{t,l}\}_{l=0}^{L_r-1}$, $\{\mathbf{g}_{r,l}\}_{l=0}^{L_r-1}$, $\{\mathbf{B}_{\hat{b}}\}_{\hat{b}=0}^{B-1}$, $\mathbf{G}_{n,m}$, $\mathbf{E}_{n,m}$, $\mathbf{G}_{b}$, $\mathbf{G}_{e}$, $\mathbf{H}_I$, $\mathbf{T}$, $\mathbf{F}$.
		\ENSURE $\mathbf{W}_c$, $\mathbf{B}$, $\mathbf{T}$, $\mathbf{F}$, $\mathbf{u}_b$
		\STATE Initialize $\mathbf{F}$, $\mathbf{B}$, $\mathbf{u}_b$, $\alpha$.
		\STATE Select the position and rotation that maximize the transmission rate using \eqref{eq47} or \eqref{eq48} to obtain $\mathbf{T}^*$.
		\STATE Optimize $\mathbf{W}_c^*$ and ${\mathbf{F}}$ using MRT and PA.
		\STATE Compute the phase shift matrix for the T-RIS using \eqref{eq54} and \eqref{eq55}, and optimize the phase shift matrix for the T-ARIS via LT.
		\STATE Calculate $\mathbf{u}_b$ using MMSE.
		\STATE\textbf{return} $\mathbf{W}_c^*$, $\mathbf{B}^*$, $\mathbf{T}^*$, $\mathbf{F}^*$, $\mathbf{u}_b$.
	\end{algorithmic}
\end{algorithm}

\subsection{Sensing and Maximization of SR}
With $\mathbf{W}_c^*$, $\mathbf{T}^*$, and $\mathbf{F}^*$, Eve can be sensed at Rx and AN is applied for jamming. Then, the corresponding subproblem based on P1 can be given by
\begin{subequations}
	\begin{align}
		\label{67}
		\text{P8:}\quad&\underset{\mathbf{W}_r,\mathbf{W}_a,\mathbf{u}_e,\alpha,\varrho_0}{\max}\ \frac{1+A_1(A_4+J_2)^{-1}}{1+A_3(A_2+J_3+\sigma_e^2)^{-1}}
		\\ 
		&\quad\mathrm{s.t.}\  \text{C1}^{\prime\prime\prime}:\text{Tr}(\mathbf{W}_{r}^H\mathbf{F}^*\mathbf{W}_{r})=\varrho_0\text{Tr}(\mathbf{W}_{c}^H\mathbf{F}^*\mathbf{W}_{c}), \\&\quad\quad\ \ \text{C2},\text{C4},\text{C5},\text{C6},
	\end{align}
\end{subequations}
where $A_1=\|t_{n,m}^*[\mathbf{Y}_{b,c}]_{n,m}\|^2$, $A_2=\|t_{n,m}^*[\mathbf{Y}_{e,r}]_{n,m}\|^2$, $A_3=\|t_{n,m}^*[\mathbf{Y}_{e,c}]_{n,m}\|^2$, and $A_4=\|t_{n,m}^*[\mathbf{Y}_{b,r}]_{n,m}\|^2$. $0\leq\varrho_0\leq1$ symbolizes the power allocation between communication and sensing at the BS. Using null-space projection, we have $\mathbf{W}_a=\mathbf{P}_a\hat{\mathbf{W}}_a$, where 
\begin{align}
	\mathbf{P}_a=\mathbf{I}_{A_t}-\mathbf{G}_{b}\mathbf{B}(\mathbf{B}^H\mathbf{G}_{b}^H\mathbf{G}_{b}\mathbf{B})^{-1}\mathbf{B}^H\mathbf{G}_{b}^H
\end{align}
denotes the NP matrix. Then, P8 can be expressed as
\begin{subequations}\label{69}
	\begin{align}
		&\text{P8:}\notag\\ &\underset{\mathbf{W}_r,\hat{\mathbf{W}}_a,\mathbf{u}_e,\alpha,\varrho_0}{\max}\ \!\!\!\frac{A_1\text{Tr}(\mathbf{W}_r^H\mathbf{A}_1\mathbf{W}_r)+A_1\text{Tr}(\hat{\mathbf{W}}_{a}^H\mathbf{A}_2\hat{\mathbf{W}}_{a})+A_1\sigma_e^2}{A_3\text{Tr}(\mathbf{W}_r^H\mathbf{A}_3\mathbf{W}_r)+A_3J_2}
		\\ 
		&\mathrm{s.t.}\  \text{C1}^{\prime\prime\prime},\text{C4},\\ &\quad\ \ \text{C2}^{\prime}:\text{Tr}(\hat{\mathbf{W}}_{a}^H\mathbf{P}_a^H\mathbf{F}\mathbf{P}_a\hat{\mathbf{W}}_{a}) = 1, \\ &\quad\ \ \text{C5}^{\prime}:\frac{\text{Tr}(\mathbf{W}_r^H\mathbf{A}_4\mathbf{W}_r)+\text{Tr}(\hat{\mathbf{W}}_{a}^H\mathbf{A}_5\hat{\mathbf{W}}_{a})}{\text{Tr}(\mathbf{W}_r^H\mathbf{A}_6\mathbf{W}_r)+\text{Tr}(\hat{\mathbf{W}}_{a}^H(\mathbf{A}_7+\mathbf{A}_8)\hat{\mathbf{W}}_{a})+J_6}\notag\\&\qquad\qquad\geq \Gamma_r,
	\end{align}
\end{subequations}
where
	\begin{align}
	\mathbf{A}_1&=\alpha P_tt_{n,m}^*\mathbf{F}^*\mathbf{E}_{n,m}^*\mathbf{u}_e\mathbf{u}_e^H(\mathbf{E}_{n,m}^*)^H\mathbf{F}^*,\\
		\mathbf{A}_2&=(1-\alpha)P_t\mathbf{P}_a^H\mathbf{G}_e\mathbf{u}_e\mathbf{u}_e^H\mathbf{G}_e^H\mathbf{P}_a,\\
			\mathbf{A}_3&=\alpha P_tt_{n,m}^*\mathbf{F}^*\mathbf{G}_{n,m}^*\mathbf{B}\mathbf{u}_b\mathbf{u}_b^H(\mathbf{B})^H(\mathbf{G}_{n,m}^*)^H\mathbf{F}^*,\\
	\mathbf{A}_4&=t_{n,m}^*\rho_0^2{\alpha P_t}\mathbf{F}^*\mathbf{g}_{n,m,0}^*\mathbf{g}_{r,0}^H\mathbf{g}_{r,0}(\mathbf{g}_{n,m,0}^*)^H\mathbf{F}^*,\end{align}
		\begin{align}
		\mathbf{A}_5&=\rho_0^2{(1-\alpha) P_t}\mathbf{P}_a^H\mathbf{F}^*\mathbf{g}_{t,0}\mathbf{g}_{r,0}^H\mathbf{g}_{r,0}\mathbf{g}_{t,0}^H\mathbf{F}^*\mathbf{P}_a,\\
	\mathbf{A}_6&={\alpha P_t}t_{n,m}^*(\sum_{l=1}^{L_r-1}\!\!\rho_l\mathbf{g}_{r,l}(\mathbf{g}_{n,m,l}^*)^H\mathbf{F}^*)^H\!\!\!\sum_{l=1}^{L_r-1}\!\!\rho_l\mathbf{g}_{r,l}(\mathbf{g}_{n,m,l}^*)^H\mathbf{F}^*,\\
		\mathbf{A}_7&={(1-\alpha) P_t}(\sum_{l=1}^{L_r-1}\rho_l\mathbf{g}_{r,l}\mathbf{g}_{t,l}^H\mathbf{F}^*\mathbf{P}_a)^H\sum_{l=1}^{L_r-1}\rho_l\mathbf{g}_{r,l}\mathbf{g}_{t,l}^H\mathbf{F}^*\mathbf{P}_a,\\
		\mathbf{A}_8&=g_I\mathbf{P}_a^H\mathbf{H}_I^H{\mathbf{H}_I\mathbf{P}_a}.
\end{align}
Correspondingly, (69d) can be rewritten as 
\begin{align}\label{78}
\text{C5}^{\prime\prime}:&\text{Tr}({\mathbf{W}_r^H(\mathbf{A}_4-\Gamma_r\mathbf{A}_6)\mathbf{W}_r})\geq\notag\\& \text{Tr}({\hat{\mathbf{W}}_{a}^H(\Gamma_r(\mathbf{A}_7+\mathbf{A}_8)-\mathbf{A}_5)\hat{\mathbf{W}}_{a}})+\Gamma_rJ_6.
\end{align}
 P8 admits a solution via the traditional Dinkelbach method but incurs high complexity~\cite{Zou2024}. To address this, closed-form solutions are provided to achieve complexity reduction. The key idea is to derive the overall solution from the solutions of two special cases. Specifically, P8 is decomposed into two scenarios: one where only the BS transmits the sensing signal, and another where only the Rx transmits the sensing signal. The optimal solution is then obtained through power allocation.
 
 In the special scenarios, the Rx operates in receive-only mode (no AN), while the BS transmits both communication and sensing signals, with the sensing signal jamming the Eve. According to \eqref{69} and \eqref{78}, the optimization problem can be formulated as
 \begin{subequations}\label{79}
 	\begin{align}
 		\text{P9:}\quad&\underset{\mathbf{W}_r,\mathbf{u}_e,\alpha,\varrho_0}{\max}\ \frac{A_1\text{Tr}(\mathbf{W}_r^H\mathbf{A}_1\mathbf{W}_r)+A_1\sigma_e^2}{A_3\text{Tr}(\mathbf{W}_r^H\mathbf{A}_3\mathbf{W}_r)+A_3J_2}
 		\\ 
 		&\mathrm{s.t.}\  \text{C1}^{\prime\prime\prime},\text{C4},\\ &\quad\ \  \text{C5}^{\prime\prime\prime}:\text{Tr}({\mathbf{W}_r^H(\mathbf{A}_4-\Gamma_r\mathbf{A}_6)\mathbf{W}_r})\geq\Gamma_rJ_6.
 	\end{align}
 \end{subequations}
By fixing $\mathbf{u}_e$, $\alpha$, and $\varrho_0$, the subproblem w.r.t. $\mathbf{W}_r$ can be formulated as
  \begin{subequations}\label{80}
 	\begin{align}
 		\text{P10.1:}\quad&\underset{\mathbf{W}_r}{\max}\ \frac{\text{Tr}(\mathbf{W}_r^H(\mathbf{A}_1+\mathbf{A}_4-\Gamma_r\mathbf{A}_6)\mathbf{W}_r)}{\text{Tr}(\mathbf{W}_r^H\mathbf{A}_3\mathbf{W}_r)}
 		\\ 
 		&\mathrm{s.t.}\  \text{C1}^{\prime\prime\prime}.
 	\end{align}
 \end{subequations}
 P10 can be solved by computing the eigenvectors corresponding to the $\Upsilon_r$ largest eigenvalues of $\mathbf{A}_3^{-1}(\mathbf{W}_r^H(\mathbf{A}_1+\mathbf{A}_4-\Gamma_r\mathbf{A}_6)$.
 
 Accordingly, the task of sensing and jamming Eve is primarily performed by the Rx, and the subproblem w.r.t. $\hat{\mathbf{W}}_a$ can be formulated as
   \begin{subequations}\label{81}
 	\begin{align}
 		\text{P10.2:}\quad&\underset{\hat{\mathbf{W}}_a}{\max}\ \frac{\text{Tr}(\hat{\mathbf{W}}_a^H\mathbf{A}_2\hat{\mathbf{W}}_a)}{\text{Tr}(\hat{\mathbf{W}}_a^H(\Gamma_r(\mathbf{A}_7+\mathbf{A}_8)-\mathbf{A}_5)\hat{\mathbf{W}}_a)}
 		\\ 
 		&\mathrm{s.t.}\  \text{C2}^{\prime}.
 	\end{align}
 \end{subequations}
Similarly, $\hat{\mathbf{W}}_a^*$ can be obtained by computing the eigenvectors corresponding to the top $\Upsilon_a$ largest eigenvalues of $(\Gamma_r(\mathbf{A}_7+\mathbf{A}_8)-\mathbf{A}_5)^{-1}\mathbf{A}_2$. Then, $\mathbf{W}_a^*=\mathbf{P}_a\hat{\mathbf{W}}_a^*$ can be calculated.

With $\mathbf{W}_r^*$ and $\mathbf{A}_a^*$ obtained, $\alpha$ and $\varrho_0$ can be subsequently determined. According to \eqref{69}, the subproblem w.r.t. $\alpha$ and $\varrho_0$ can be expressed as
\begin{subequations}\label{82}
	\begin{align}
		&\text{P11:}\notag\\ &\underset{\alpha,\varrho_0}{\max}\quad \frac{1+\alpha  (1-\varrho_0)\hat{A}_1(\alpha  \varrho_0D_1-\alpha D_2+D_2+\sigma_e^2)}{1+\alpha  (1-\varrho_0)\hat{A}_3(\alpha\varrho_0 D_3+J_2)}\label{82a}
		\\ 
		&\mathrm{s.t.}\  \text{C10}:\frac{\alpha\varrho_0 D_4-\alpha D_5+D_5}{\alpha\varrho_0 D_6-\alpha D_7+D_7+D_8+\alpha(1-\varrho_0) J_6}\geq \Gamma_r,
		%&\quad\ \ \text{C10}:
	\end{align}
\end{subequations}
 where
 	\begin{align}
 		\hat{A}_1&=\alpha^{-1}{A}_1, \hat{A}_3=\alpha^{-1}{A}_3,\\
D_1 &= \text{Tr}(\alpha^{-1}(\mathbf{W}_r^*)^H\mathbf{A}_1\mathbf{W}_r^*),\\
D_2 &= \text{Tr}(P_t(\mathbf{W}_a^*)^H\mathbf{P}_a^H\mathbf{G}_e\mathbf{u}_e\mathbf{u}_e^H\mathbf{G}_e^H\mathbf{P}_a\mathbf{W}_a^*),\\
D_3 &= \text{Tr}(\alpha^{-1}(\mathbf{W}_r^*)^H\mathbf{A}_3\mathbf{W}_r^*),\\
D_4 &= \text{Tr}(\alpha^{-1}(\mathbf{W}_r^*)^H\mathbf{A}_4\mathbf{W}_r^*),\\
D_5 &= \text{Tr}(\rho_0^2P_t(\mathbf{W}_a^*)^H\mathbf{F}^*\mathbf{g}_{t,0}\mathbf{g}_{r,0}^H\mathbf{g}_{r,0}\mathbf{g}_{t,0}^H\mathbf{F}^*\mathbf{W}_a^*),\\
% \end{align}
%\begin{align}
 	D_6 &=\text{Tr}(\alpha^{-1}{ P_t}(\mathbf{W}_r^*)^H\mathbf{A}_6\mathbf{W}_r^*),\\
 	D_7 &= \text{Tr}((\sum_{l=1}^{L_r-1}\rho_l\mathbf{g}_{r,l}\mathbf{g}_{t,l}^H\mathbf{F}^*\mathbf{W}_a^*)^H\!\sum_{l=1}^{L_r-1}\rho_l\mathbf{g}_{r,l}\mathbf{g}_{t,l}^H\mathbf{F}^*\mathbf{W}_a^*),\\
 	D_8 &=\text{Tr}(g_I(\mathbf{W}_a^*)^H\mathbf{H}_I^H{\mathbf{H}_I\mathbf{W}_a^*}).
 \end{align}
 The equality condition of constraint C10 yields 
 \begin{align}\label{92}
 \alpha &=( \Gamma_r(D_7+D_8)-D_5)\notag\\&\quad\ (\rho_0(D_4-\Gamma_rD_6+J_6)-D_5-\Gamma_rD_7-J_6)^{-1}. 
\end{align}
 Then, substituting \eqref{92} into \eqref{82a}, $\rho_0^*$ and $\alpha^*$ can be calculated.
Similarly, with $\mathbf{W}_r^*$, $\mathbf{W}_a^*$, $\rho_0^*$, and $\alpha^*$, $\mathbf{u}_e^*$ can be derived using the MMSE. The overall two-stage algorithm is illustrated in Algorithm \ref{alg2}.
 \begin{algorithm}[t]
 	\renewcommand{\algorithmicrequire}{\textbf{Input:}}
 	\renewcommand{\algorithmicensure}{\textbf{Output:}}
 	\caption{Two-stage Optimization Algorithm.}\label{alg2}
 	\begin{algorithmic}[1]
 		\REQUIRE $P_t$, $\Gamma_r$, $d$, $P_{\text{RIS}}$, $\Upsilon_c$, $\Upsilon_r$, $\Upsilon_a$, $\theta_{\text{3dB}}$, $\phi_{\text{3dB}}$, $\sigma_b$, $\sigma_e$, $\sigma_r$, $\{\varsigma_{n,\iota}\}_{\iota=0}^{L_b-1}$, $\{\rho_l\}_{l=0}^{L_r-1}$, $\{\mathbf{g}_{n,m,l}\}_{l=0}^{L_r-1}$, $\{\mathbf{g}_{t,l}\}_{l=0}^{L_r-1}$, $\{\mathbf{g}_{r,l}\}_{l=0}^{L_r-1}$, $\{\mathbf{B}_{\hat{b}}\}_{\hat{b}=0}^{B-1}$, $\mathbf{G}_{n,m}$, $\mathbf{E}_{n,m}$, $\mathbf{G}_{b}$, $\mathbf{G}_{e}$, $\mathbf{H}_I$, $\mathbf{T}$, $\mathbf{F}$.
 		\ENSURE $\mathbf{W}$, $\alpha$, $\mathbf{B}$, $\mathbf{T}$, $\mathbf{F}$, $\mathbf{u}$
 		\STATE Initialize $\mathbf{F}$, $\mathbf{B}$, $\mathbf{u}_b$, $\alpha$.
 		\STATE Optimize $\mathbf{T}^*$ by solving \eqref{eq47} or \eqref{eq48} through the GRQ method.
 		\STATE Optimize $\mathbf{W}_c^*$ and ${\mathbf{F}}$ using \eqref{51} and \eqref{52}.
 		\STATE Optimize $\mathbf{W}_r^*$ and $\mathbf{W}_a^*$ using \eqref{80} and \eqref{81}.
 		\STATE Compute $\alpha^*$ by solving \eqref{82}.
 		\STATE Optimize $\mathbf{B}^*$ by solving \eqref{eq54}, \eqref{eq55}, and \eqref{57}.
 		\STATE Compute $\mathbf{u}_b$ and $\mathbf{u}_e$ using MMSE.
 		\STATE\textbf{return} $\mathbf{W}_c^*$, $\mathbf{W}_r^*$, $\mathbf{W}_a^*$, $\alpha^*$, $\mathbf{B}^*$, $\mathbf{T}^*$, $\mathbf{F}^*$, $\mathbf{u}_b$, $\mathbf{u}_e$.
 	\end{algorithmic}
 \end{algorithm}

\section{AI-enabled Joint Optimization Scheme}\label{s4}
\subsection{Motivation of Adopting MADDPG}
\begin{figure}
	\centering
	\includegraphics[width=0.40\textwidth, trim = 20 20 10 10,clip]{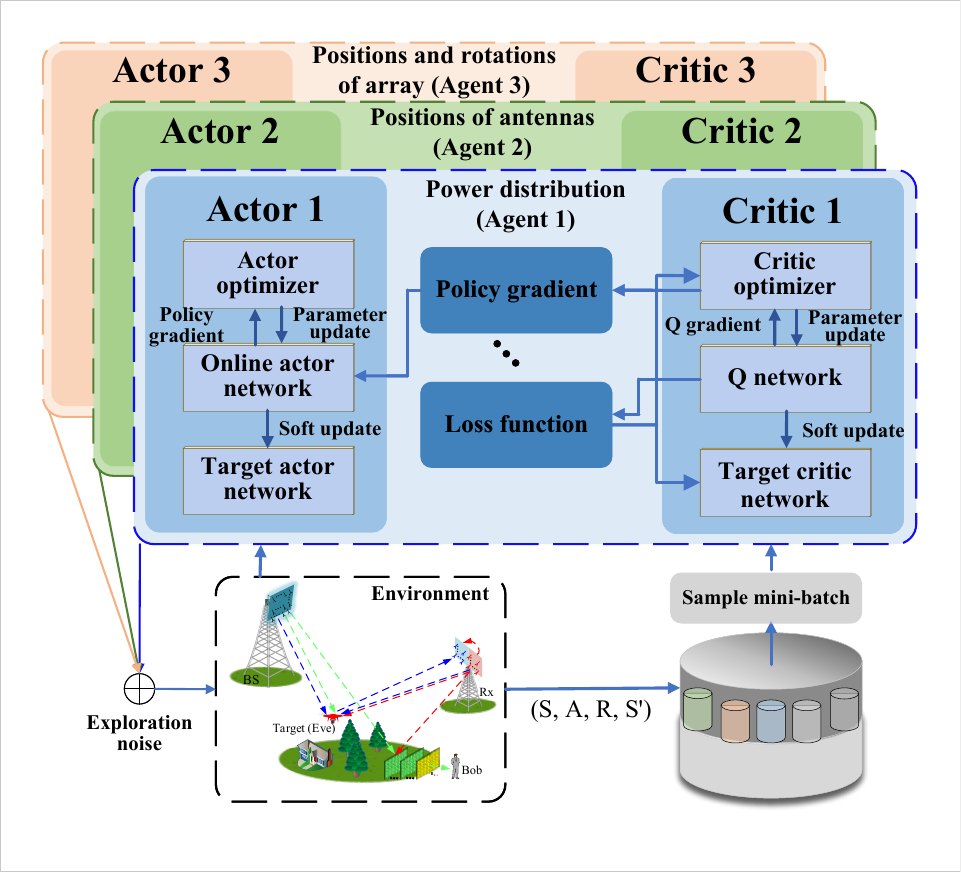}\\
	\caption{The illustration of MADDPG-based SR optimization approach.}\label{f3}
\end{figure}
The proposed two-stage algorithm is based on a branch-and-bound approach, which is prone to converging to local optima for the objective function. Furthermore, relaxation of non-convex problems typically introduces inherent performance losses, thus limiting the achievable system performance. To achieve enhanced security, a SR optimization framework based on the Multi-Agent Deep Deterministic Policy Gradient (MADDPG) is proposed. Additionally, closed-form solutions are introduced to reduce the complexity of the network structure.
\subsection{Markov Decision Process Modeling}
The optimization problem can be modeled as a multi-agent Markov decision process (MDP), given by $(S_i^j,A_i^j,R_i^j,S_{i+1}^j,j)$. $S_i^j$, $A_i^j$, $R_i^j$, and $S_{i+1}^j$ denote  the observation, action, reward, and next observation for the $j$-th ($j=1,2,3$) agent, respectively~\cite{Zheng2025}. To enhance the security of medium-to-long-range communications with an Eve on the line-of-sight path, the agents learn to optimize the beam pattern by exploring optimal antenna positions and power allocation, and adjust the array orientation through optimal positioning and rotation. The key elements of MADDPG are described as follows in this work.
\subsubsection{Agents}
Three agents are employed, tasked with optimizing the array pose (position and rotation), the antenna positions, and the power allocation strategy, respectively.
\subsubsection{Observations}
For the $t$-th time slot, the observations of the three agents are denoted by
\begin{align}
	S_i^1 &= \{r_b,r_e,R_s,t_{n,m},p_{\mathbf{F}}\}, \\S_i^2 &= \{r_b,r_e,R_s,t_{n,m},\alpha,\varrho_0\},\\S_i^3 &= \{r_b,r_e,R_s,p_{\mathbf{F}},\alpha,\varrho_0\},
\end{align}
respectively. Here, $p_{\mathbf{F}}$ is obtained by vectorizing matrix $\mathbf{F}$, converting it to a binary representation, and then transforming it into a decimal number.
\subsubsection{Actions}
The action space in MADDPG encompasses the $NM$ array poses (composed of $N$ rotations and $M$ positions), the locations of $A$ antennas, along with the continuous actions $\alpha$ and $\varrho_0$.
\subsubsection{Rewards}
Following P1, the reward $R_s$ is designed to incentivize security performance, whereby a higher reward corresponds to a better action.
\subsubsection{Replay Buffer}
The replay buffer is used for storing the three agents' observation vectors for the current and next states, their actions, and rewards, from which minibatches are provided to the actor and critic for training.
\subsection{SR Optimization Algorithm Based on MADDPG}
The SR optimization approach based on MADDPG is illustrated in Fig. \ref{f3}. The $j$-th agent's policy and value functions are represented by actor $j$ and critic $j$, respectively. They are used for action selection and $Q$-value estimation, where the $Q$-value denotes the expected cumulative future reward after taking action $A_i^j$ in state $S_i^j$. Target networks are established to ensure training stability.
First, the actor and critic networks for the three agents are initialized. Subsequently, each agent generates actions based on its current policy and interacts with the environment, yielding new observations and rewards. The experiences gained from these interactions are stored in the replay buffer. Once sufficient experience is accumulated, random mini-batches of data are sampled from the buffer for training iterations.

Specifically, each online actor network outputs an action $A_i^j$ based on its current local observation $S_i^j$. To facilitate exploration, Gaussian noise is added to $A_i^j$. The actions from all agents collectively form the action set $A_i=\{A_i^1,A_i^2,A_i^3\}$. By inputting $A_i$ into the environment and executing Algorithm \ref{alg2}, a new observation set $S_i=\{S_{i+1}^1,S_{i+1}^2,S_{i+1}^3\}$ and a reward set $R_i=\{R_i^1,R_i^2,R_i^3\}$ can be obtained. The experience data from all agents is stored in the replay buffer $D$. When a sufficient number of $I_D$ experience tuples are accumulated, a mini-batch of size $\varpi$ is sampled for training.
For each transition in the mini-batch, the next observation is fed into the target actor network to compute the next action, yielding the set of next actions for all sampled experiences. This set of next actions and the set of next observations are then input into the target critic network to compute the target $Q$-value, i.e., $Q_{\theta_Q}(S_{i}, A_{i})$.  The $Q$-network is subsequently updated by minimizing the temporal difference error, and its parameters $\theta_{Q}=\{\theta_{Q}^{1},\theta_{Q}^{2},\theta_{Q}^{3}\}$ are refined via gradient descent. The loss function and gradient formula can be expressed by
\begin{align}
	\min_{\theta_{Q}^{j}} \mathrm{loss} &= \min_{\theta_{Q}^{j}} \mathbb{E}_{(S_i^j,A_i^j,R_i^j,S_{i+1}^j,j) \sim D} \Bigl[\Bigl(Q_{\theta_Q}(S_i,A_i)\notag\\
	&\quad-(R_i^j+\gamma Q_{\theta_Q}(S_{i+1},A_{i+1}))\Bigr)^2\Bigr],\label{96}\\
	\nabla_{\theta_{Q}^{j}}J(\theta_{Q}^{j}) & =\nabla_{\theta_{Q}^{j}}\mathbb{E}_{(S_i^j,A_i^j,R_i^j,S_{i+1}^j,j) \sim D}\Bigl[\Bigl(Q_{\theta_Q}(S_i,A_i)\notag\\
	&\quad -(R_i^j+\gamma Q_{\theta_Q}(S_{i+1},A_{i+1}))\Bigr)^2\Bigr],\label{97}
\end{align}
where $\gamma$ denotes the learning rate. Following this, the actor networks are optimized using the policy gradient method, which aims to maximize the $Q$-value by selecting the optimal action given a specific observation.
After updating the online networks, the soft update is applied to the target networks. The iterative process concludes once the rewards stabilize or the maximum number $\Lambda$ of training episodes is reached. The overall MADDPG-based SR optimization algorithm is illustrated in Algorithm \ref{alg3}.
\begin{algorithm}[t]
	\renewcommand{\algorithmicrequire}{\textbf{Input:}}
	\renewcommand{\algorithmicensure}{\textbf{Output:}}
	\caption{MADDPG-based SR Optimization Algorithm}
	\label{alg3}
	\begin{algorithmic}[1]
		\REQUIRE $P_t$, $\Gamma_r$, $d$, $P_{\text{RIS}}$, $\Upsilon_c$, $\Upsilon_r$, $\Upsilon_a$, $\theta_{\text{3dB}}$, $\phi_{\text{3dB}}$, $\sigma_b$, $\sigma_e$, $\sigma_r$, $\{\varsigma_{n,\iota}\}_{\iota=0}^{L_b-1}$, $\{\rho_l\}_{l=0}^{L_r-1}$, $\{\mathbf{g}_{n,m,l}\}_{l=0}^{L_r-1}$, $\{\mathbf{g}_{t,l}\}_{l=0}^{L_r-1}$, $\{\mathbf{g}_{r,l}\}_{l=0}^{L_r-1}$, $\{\mathbf{B}_{\hat{b}}\}_{\hat{b}=0}^{B-1}$, $\mathbf{G}_{n,m}$, $\mathbf{E}_{n,m}$, $\mathbf{G}_{b}$, $\mathbf{G}_{e}$, $\mathbf{H}_I$, $D$, $\varpi$, $\Lambda$.
		\ENSURE $\alpha^*$, $\mathbf{T}^*$, $\mathbf{F}^*$
		\STATE Initialize actor and critic networks for the three agents.
		\FOR{episode = 1 to $\Lambda$}
		\FOR{$i$ = 1 to $I_s$}
		\STATE Select the optimal action for each agent based on its local observation.
		\STATE Execute the actions in the environment, acquiring rewards via Algorithm \ref{alg2}.
		\STATE Store the experience tuple $(S_i^j,A_i^j,R_i^j,S_{i+1}^j)$ in the replay buffer $D$. Then, set $i_D = i$.
		\IF{$i_D\geq I_D$}
		%\FOR{episode = 1 to $\Lambda$}
		\STATE Sample a mini-batch of size $\varpi$ randomly from $D$.
		\STATE Compute the target $Q$-value $Q_{\theta_Q}(S_{i}, A_{i})$.
		\STATE Update the $Q$-network and its parameters according to \eqref{96} and \eqref{97}.
		\STATE Update the online actor network using the temporal difference error and gradient descent.
		\ENDIF
		\STATE Update all target network parameters via soft update.
		\ENDFOR
		\ENDFOR
		\RETURN $\alpha^*$, $\mathbf{T}^*$, $\mathbf{F}^*$.
	\end{algorithmic}
\end{algorithm}

\section{Simulation Results}\label{s5}
\begin{table}[t]
	\centering\renewcommand{\arraystretch}{1.2}
	\setlength{\tabcolsep}{19pt}
	\caption{Parameters of MADDOG}\label{tab}
	\begin{tabular}{l|c}
		%\rowcolor{black!20}\hline
		\toprule
		\textbf{Parameter}& \textbf{Value}  \\ \bottomrule
		Learning rate of actor network & 0.01 \\ \hline
		Learning rate of critic network& 0.002 \\ \hline
		Batch size & 128\\\hline 
		The rate of soft update & 0.001 \\ \hline
		Experience replay buffer size & 5000 \\ \hline
		Exploration noise & Gaussian noise \\ \hline
		%\bottomrule
	\end{tabular}
\end{table}
In this section, simulation results are provided. The number of antennas at the BS, the number of elements per RIS, and the number of transmit/receive antennas at the Rx are set to $A=10$, $N_R=64$, and $A_t=A_r=100$, respectively. Bob and Eve are each equipped with a single antenna.
The numbers of candidate positions and rotations are $M=225$ and $N=512$, respectively. The number of discrete positions on the RA array is $\hat{Q}=100$. We consider $3$ effective channel paths, with a path gain of $-20.03$ dB per meter. The noise levels are $\sigma_e^2=\sigma_b^2=-80$ dBm and $\sigma_I^2=\sigma_r^2=-90$ dBm, respectively. Unless otherwise specified, the carrier frequency, transmit power, and maximum power for T-ARIS are set to $2.4$ GHz, $40$ dBm, and $-10$ dBm, respectively~\cite{Zheng2025a}. The distances from the BS to the Rx, from the BS to the T-RIS, and from the Rx to the T-RIS are $2000$ m, $3000$ m, and $1500$ m, respectively. The coordinate of Eve is $[10,2900,10]^T$. The distance between adjacent RIS layers is set to $0.03$ m. The key MADDPG parameters are listed in Table \ref{tab}~\cite{Zheng2025}. For each antenna element, we consider the following radiation pattern~\cite{Balanis2016}:
\begin{equation}
	 G_a(\theta^{\prime}_a,\phi^{\prime}_a) = \begin{cases}
	 	G_{a,0}\cos^{2p}(\phi^{\prime}_a) &  \theta^{\prime}_a\in[0,\tfrac{\pi}{2}),\phi^{\prime}_a\in[0,2{\pi}) ,\\
	 	0  &  \text{otherwise},
	 \end{cases}
\end{equation}
where $p$ denotes the antenna directivity factor that symbolizes the mainlobe beamwidth, $G_{a,0}=2(2p+1)$ is the maximum gain, and $(\theta^{\prime}_a,\phi^{\prime}_a)$ denotes a pair of elevation and azimuth angles in the local coordinate system of the antenna element. 
\begin{figure}[t]
	\centering
	\includegraphics[width=0.40\textwidth, trim = 20 1 20 20,clip]{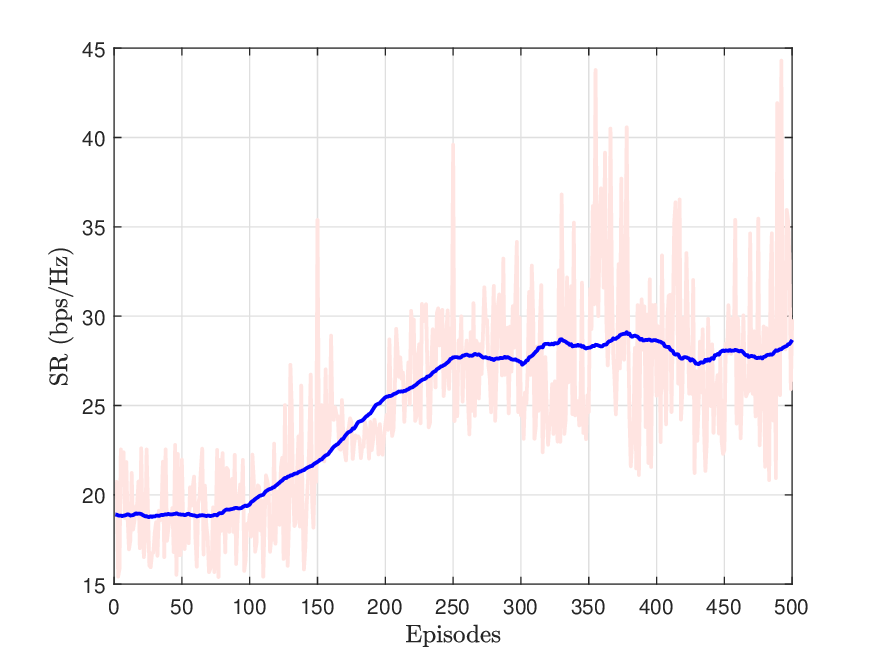}\\
	\caption{Convergence of the proposed Algorithm \ref{alg3}.}\label{fig4}
\end{figure}
\begin{figure}[htp]
	\centering
	%\begin{subfigures}
	\subfloat[]{\label{fig5a}
		\includegraphics[width=1.6in, trim = 3 3 30 20,clip]{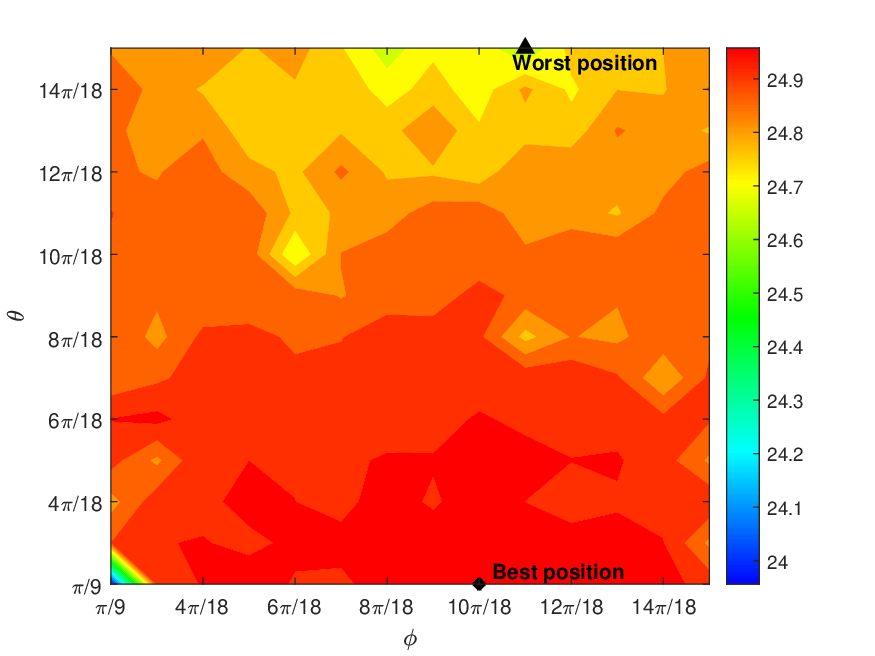}}\hfil
	\subfloat[]{\label{fig5b}
		\includegraphics[width=1.6in, trim = 3 3 30 20,clip]{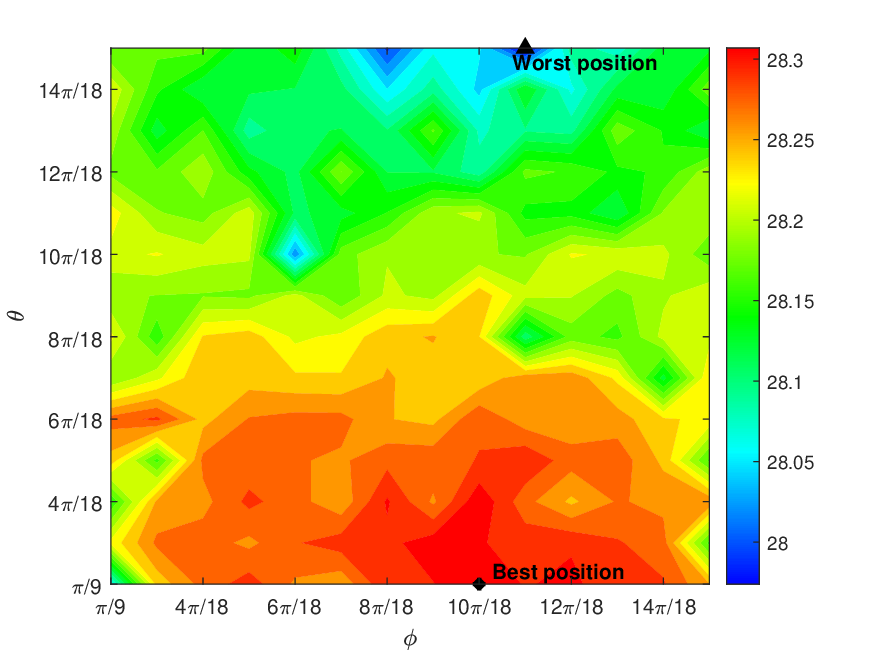}}\\
	\subfloat[]{\label{fig5c}
		\includegraphics[width=1.6in, trim = 3 3 30 20,clip]{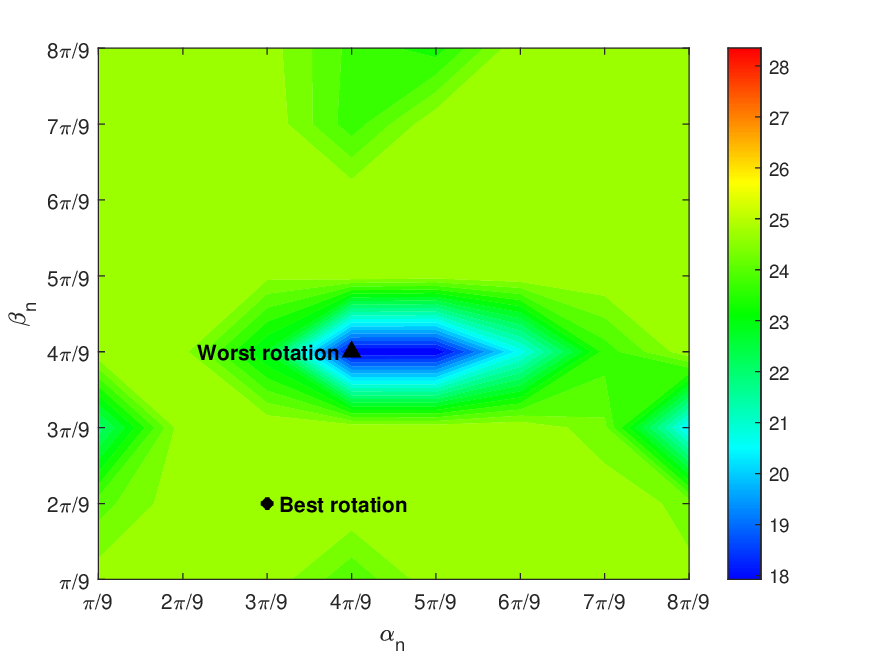}}\hfil
	\subfloat[]{\label{fig5d}
		\includegraphics[width=1.6in, trim = 3 3 30 20,clip]{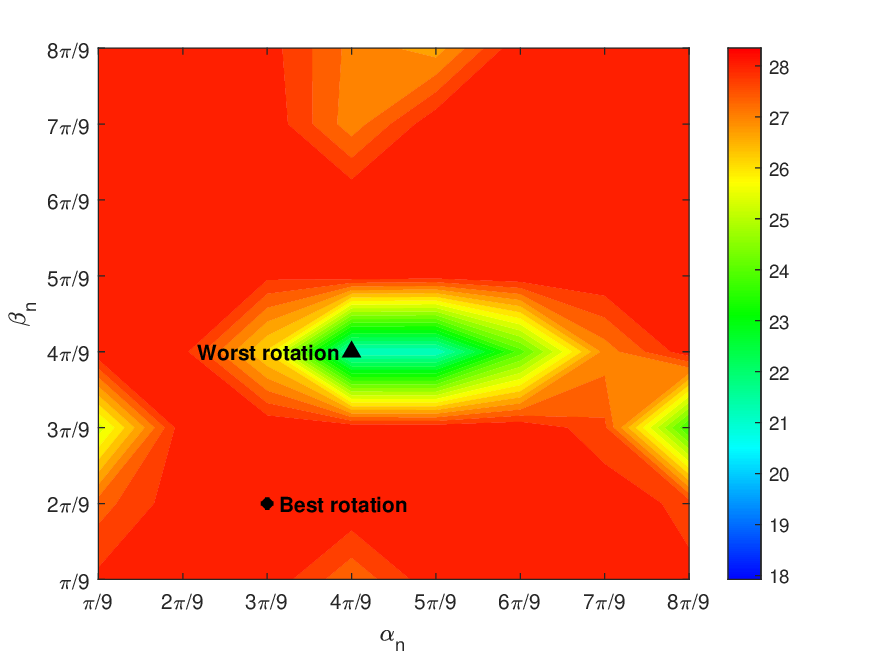}}\\
	\caption{SR under different positions and rotations. (a) SR versus position before T-RIS optimization. (b) SR versus position after T-RIS optimization. (c) SR versus rotation before T-RIS optimization. (d) SR versus rotation after T-RIS optimization.}\label{fig5}	%\end{subfigures}
\end{figure}

The convergence behavior of Algorithm \ref{alg3} is illustrated in Fig. \ref{fig4}. As the number of episodes increases, the exploration of improved joint actions enables higher cumulative rewards. However, the learning-based algorithms generally entail higher computational complexity compared to conventional convex optimization methods, making them more suitable for offline optimization based on statistical data. In contrast, the computational overhead of Algorithm \ref{alg2} stems primarily from eigenvector computation and eigenvalue comparison, resulting in lower complexity that is practical for online optimization.

The SR under different positions and rotations is illustrated in Fig. \ref{fig5}. It can be observed that deploying the T-RIS enhances the SR performance across RA array configurations, which is primarily attributed to the signal enhancement capability of the T-RIS. Furthermore, adjusting the array rotation yields greater performance improvement compared to altering its position. Since continuous ranges of rotations and positions exhibit similar performance, a certain delay in array pose adjustment is tolerable in practical implementations. The optimization of array pose adjustment paths will be investigated in future work.
\begin{figure}[t]
	\centering
	\includegraphics[width=0.41\textwidth, trim = 20 1 20 20,clip]{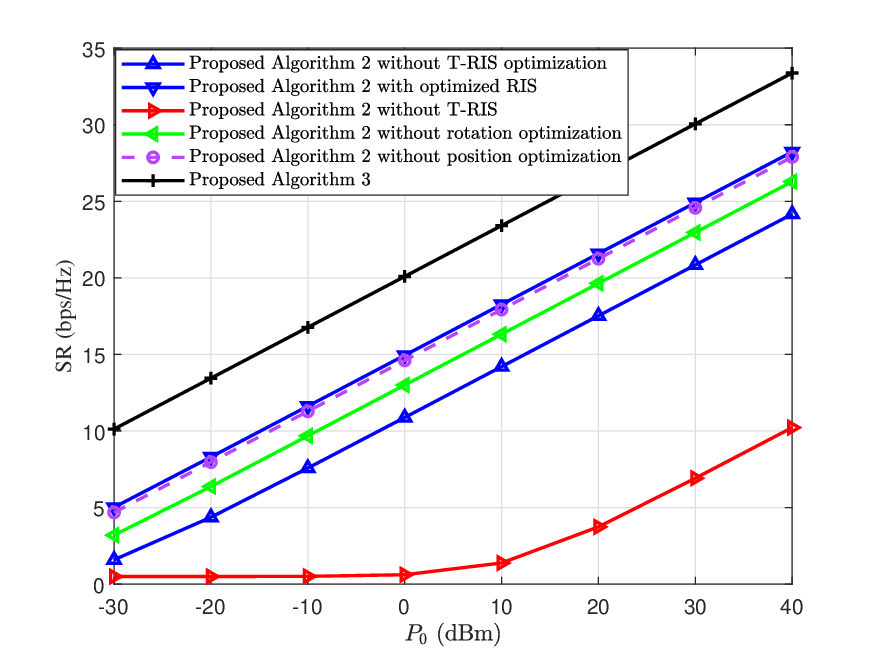}\\
	\caption{SR versus transmit power $P_0$.}\label{Fig6}
\end{figure}
\begin{figure}[t]
	\centering
	\includegraphics[width=0.41\textwidth, trim = 20 1 20 20,clip]{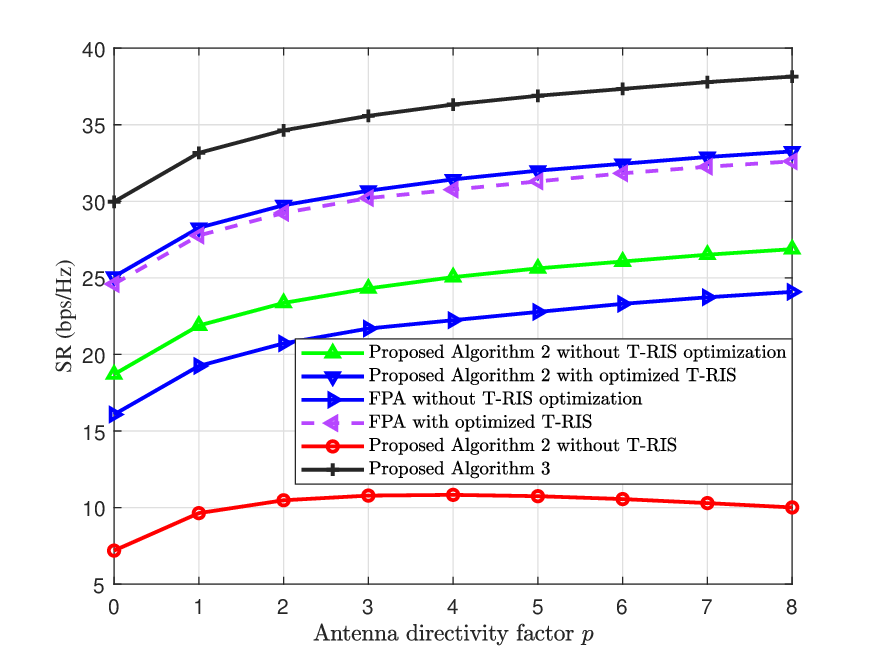}\\
	\caption{SR versus antenna directivity factor $p$.}\label{Fig7}
\end{figure}

Fig. \ref{Fig6} depicts the relationship between the SR and transmit power. The scenarios without array rotation optimization, without array position optimization, without T-RIS optimization, and without any RIS deployment are respectively adopted as baseline schemes. The security performance improves with increasing transmit power. When the transmit power is below $10$ dBm, the SR without T-RIS is only about $1$ bps/Hz, whereas deploying T-RIS yields a twofold performance improvement at $30$ dBm. Since the sensing performance must be guaranteed to support the DOA estimation of Eve, deploying T-RIS is more conducive to achieving higher security in the case of low power. A properly optimized T-RIS enables an even higher SR. Furthermore, the performance degradation due to the lack of position optimization is less severe than that from the lack of rotation optimization. Algorithm 3 outperforms all other schemes. The low-complexity Algorithm \ref{alg2} may incur a performance loss of at least $5$ bps/Hz. In practice, the performance achievable by rotating the array may fall short of the theoretical optimum.

Fig. \ref{Fig7} illustrates the dependence of the achievable SR performance on the antenna directivity factor. The ``FPA" baseline schemes represent conventional antenna configurations with half-wavelength spacing. For systems without T-RIS, the achievable SR gradually deteriorates when the directivity factor exceeds $p>4$, owing to the Eve's location on the main communication path, which consequently yields a higher pre-estimated received SNR at Eve. In contrast, deploying and optimizing the T-RIS enhances security performance with increasing $p$, validating the effectiveness of both the algorithm and the proposed system. Notably, even a non-optimized T-RIS provides security gains that improve with $p$. Furthermore, Algorithm \ref{alg3} achieves the optimal SR performance, while Algorithm \ref{alg2} outperforms the FPA system by approximately $2.6$ bps/Hz.
\begin{figure}[t]
	\centering
	\includegraphics[width=0.41\textwidth, trim = 20 1 20 20,clip]{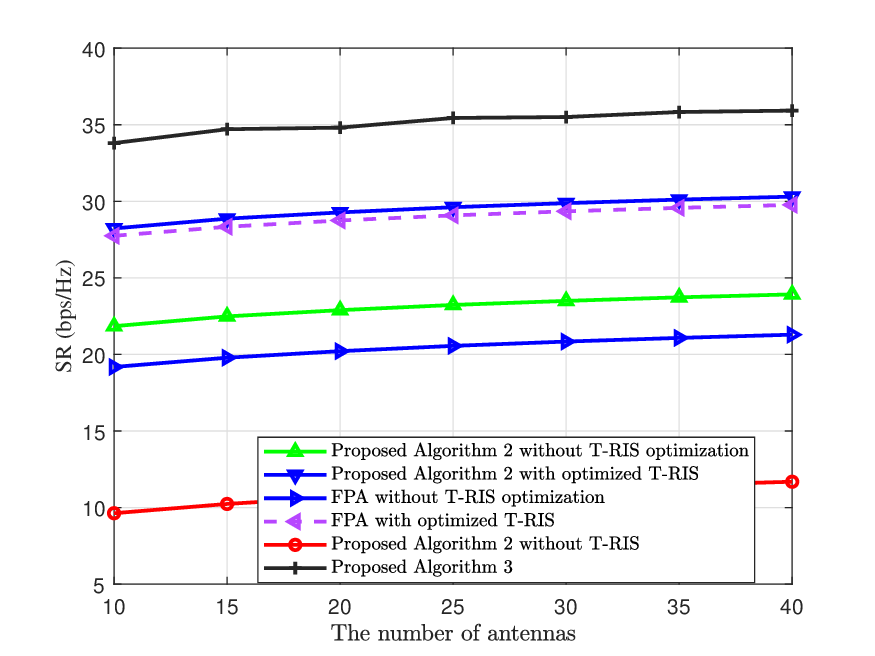}\\
	\caption{SR versus the number of antennas.}\label{Fig8}
\end{figure}
\begin{figure}[t]
	\centering
	\includegraphics[width=0.41\textwidth, trim = 20 1 20 20,clip]{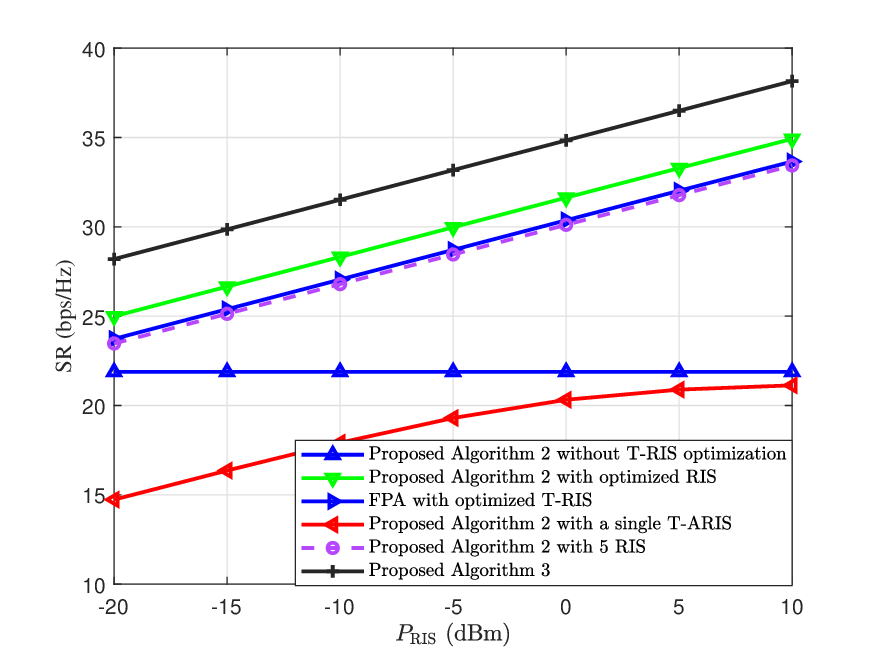}\\
	\caption{SR versus the maximum power corresponding to T-RIS.}\label{Fig10}
\end{figure}

Higher SR performance can be achievable with an increasing number of antennas, as shown in Fig. \ref{Fig8}. An incremental addition of $5$ antennas yields an SR improvement of approximately $1$ bps/Hz. Furthermore, the SR performance improves with increasing reflected power. The achievable SR with a single-layer T-ARIS is limited, saturating gradually when $P_{\mathrm{RIS}} > 5$ dBm, as the amplified noise concurrently affects the receiver. Deploying a two-layer T-PRIS elevates the SR by approximately 58\% while the interference is attenuated. However, with a five-layer RIS configuration, the SR improvement relative to the single-layer T-ARIS is about 49.6\%, which is lower than that achieved by the three-layer T-RIS. As the number of RIS layers increases further, the attainable performance gain may not compensate for the detrimental effects introduced by the multiplicative path loss.
\begin{figure}[t]
	\centering
	\includegraphics[width=0.41\textwidth, trim = 20 1 20 20,clip]{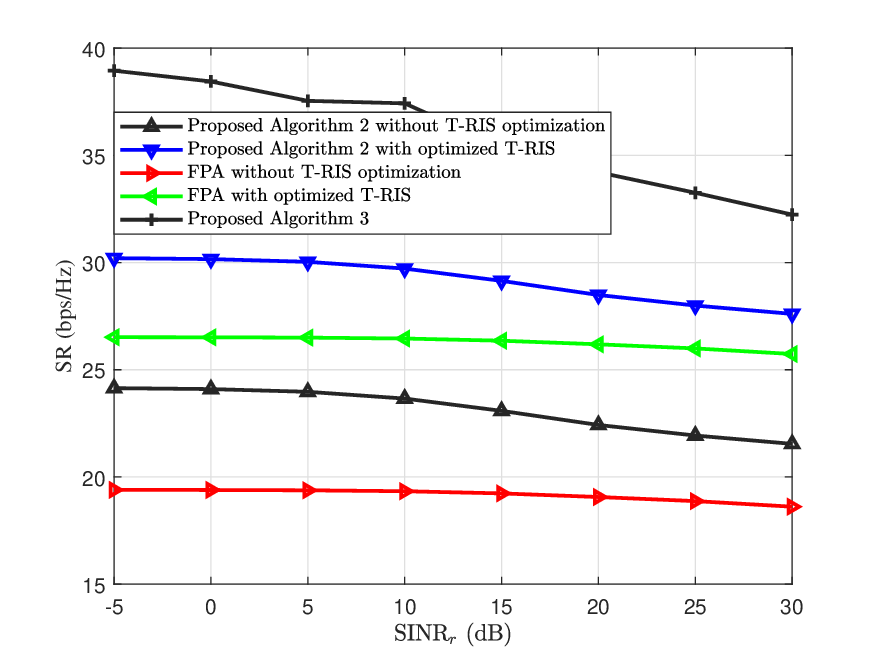}\\
	\caption{SR versus $\mathrm{SINR}_r$.}\label{Fig9}
\end{figure}

Fig. \ref{Fig9} demonstrates the trade-off between communication and sensing performance. As the sensing requirement becomes more stringent, the achievable communication performance degrades. As shown in Algorithm \ref{alg2}, after designing the precoding matrices for communication and sensing, optimizing the power allocation factor is crucial. Allocating more power to communication tasks leads to a higher transmission rate at Bob. However, since Eve is located on the main path from the BS to Bob, the estimated rate at Eve increases proportionally. Conversely, allocating more power to sensing improves sensing performance while simultaneously causing more interference to Eve. The proposed algorithms consistently outperform the conventional FPA scheme. Under the condition of $\mathrm{SINR}_r < 10$ dB, the communication performance of Algorithm \ref{alg2} with optimized T-RIS degrades minimally as the sensing requirement increases, which can be attributed to the underutilization of the potential system degrees of freedom. By exploring superior action combinations under the same sensing constraints, Algorithm \ref{alg3} achieves higher SR performance.

\section{Conclusion}\label{s6}
To address secure communication under diverse antenna radiation patterns, communication security under sensing performance constraints was investigated in this paper. To tackle the non-convex optimization and complexity challenges, two algorithms were developed: a low-complexity two-stage online algorithm based on the GRQ, and an offline algorithm based on the MADDPG. The proposed algorithms maximize the SR by jointly optimizing the rotatable array's pose, antenna distribution, T-RIS phase shift matrix, and power allocation factor. Simulation results validated the effectiveness of the proposed approaches. Compared to conventional schemes, the rotatable array architecture can improve SR performance, while the deployment of T-RIS provides further security enhancement.

%\vfill
\renewcommand\refname{References}
\bibliographystyle{IEEEtran}
\bibliography{mybib}
\end{document}